\documentclass[journal=ancac3, manuscript=article,layout=twocolumn]{achemso}

\usepackage{comment}
\usepackage{chemformula} 
\usepackage[T1]{fontenc} 
\usepackage{graphicx}
\usepackage{subcaption}
\usepackage{dcolumn}
\usepackage{tabu} 
\usepackage{mathtools}

\usepackage{braket}
\usepackage{color}
\usepackage{amsmath}
\usepackage{xcolor}
\usepackage{braket}
\usepackage[normalem]{ulem}
\usepackage{setspace}
\usepackage{booktabs}
\usepackage[inkscapelatex=false]{svg}
\usepackage{soul}
\usepackage{ragged2e}

\author{Andrew N. Wakileh}
\affiliation{Centre for Nanophotonics, Department of Physics, Engineering Physics, and Astronomy, Queen's University, Kingston, Ontario, Canada, K7L 3N6}
\alsoaffiliation{National Research Council of Canada, Ottawa, Ontario, Canada, K1A 0R6}

\author{Dan Dalacu}
\affiliation{National Research Council of Canada, Ottawa, Ontario, Canada, K1A 0R6}
\alsoaffiliation{Centre for Nanophotonics, Department of Physics, Engineering Physics, and Astronomy, Queen's University, Kingston, Ontario, Canada, K7L 3N6}

\author{Philip J. Poole}
\affiliation{National Research Council of Canada, Ottawa, Ontario, Canada, K1A 0R6}

\author{Boris Lamontagne}
\affiliation{National Research Council of Canada, Ottawa, Ontario, Canada, K1A 0R6}

\author{Simona Moisa}
\affiliation{National Research Council of Canada, Ottawa, Ontario, Canada, K1A 0R6}

\author{Robin L. Williams}
\affiliation{National Research Council of Canada, Ottawa, Ontario, Canada, K1A 0R6}

\author{Nir Rotenberg}
\affiliation{Centre for Nanophotonics, Department of Physics, Engineering Physics, and Astronomy, Queen's University, Kingston, Ontario, Canada, K7L 3N6}
\email{Nir.Rotenberg@queensu.ca}

\title{Approaching transform-limited linewidths in telecom-wavelength transitions of ungated quantum dots}


\begin{document} 

\begin{abstract}
Highly coherent quantum emitters operating in the telecommunication C-band (1530 - 1565\,nm),  where ultra-low-loss fibers and photonic circuits are available, are crucial to the development of scalable quantum technologies. In this work, we report on a modified Stranski-Krastanov growth scheme using chemical beam epitaxy to enable the generation of high-quality InAs/InP quantum dots, characterized by near-transform-limited linewidths ($\Gamma_{\mathrm{TL}}$). We demonstrate the growth of highly-symmetric quantum dots with aspect ratios >\,0.8 and densities ranging from 2 to 22\,$\mu$m$^{-2}$. Optical characterization of these sources reveal fine-structure splittings down to $25\pm4\,\mu$eV and a single-photon purity of $g^{(2)}(0) = 0.012\pm\mathrm{0.007}$, confirming the quality of these dots. Further, using an etalon to measure the linewidth, in combination with rigorous modelling, we find an upper-bound to the mean, low-power linewidths of only $12.1\pm 6.7\,\Gamma_\mathrm{TL}$ and, in the best case, $2.8\pm 1.8\,\Gamma_\mathrm{TL}$. These results represent a significant step in the development of telecom-wavelength quantum light sources which are essential for complex quantum networks and devices.
\end{abstract}

\maketitle


\section{Introduction and Background}
Highly coherent quantum emitters are crucial to the development of emerging quantum technologies. They produce indistinguishable photons over long time scales \cite{meraner_indistinguishable_2020, huber_highly_2017}, coherently scatter photons to modulate their phase or amplitude\cite{turschmann_coherent_2019, staunstrup_direct_2024}, mediate nonlinearities that enable photon-photon interactions\cite{le_jeannic_dynamical_2022, koong_coherence_2022, volz_nonlinear_2014}, or enable many-body physics\cite{pallmann_cavity-mediated_2024, singh_coherent_2025, Manzoni_Designing-Exotic_2017}. The transition linewidths, $\Gamma$, of these highly coherent emitters are ideally limited by their related radiative lifetimes $T_1$, i.e. transform-limited where $\Gamma_\mathrm{TL} = 1/{2\pi T_1}$. Reducing the transition linewidths to this limit has been one of the most important challenges of modern quantum photonics, particularly for most solid-state quantum emitters.

A second enduring challenge is the creation of telecommunication-wavelength quantum emitters that can deterministically generate and process quantum light states. In particular, emission in the C-band (between 1530\,nm and 1565\,nm), where ultra-low-loss fibres\cite{gisin_quantum_2007} and photonic circuits \cite{elshaari_hybrid_2020, alexander_manufacturable_2025} are available, has been difficult. Here, significant progress has been made recently with indium arsenide (InAs) semiconductor quantum dots (QDs) grown in gallium arsenide\cite{nawrath_resonance_2021, zeuner_-demand_2021, kaupp_purcell-enhanced_2023, joos_coherently_2024} (GaAs) or indium phosphide (InP) \cite{vajner_-demand_2024, wells_coherent_2023, holewa_droplet_2022, phillips_purcell-enhanced_2024, rahaman_efficient_2024, wakileh_single_2024} using the Stranski-Krastanov (SK) growth mode, as well as droplet epitaxy (DE), and vapour-liquid-solid (VLS) growth. Epitaxially grown InAs QDs in InP naturally emit at telecom wavelengths and, as with other platforms, are bright and have near unity quantum efficiency \cite{smolka_optical_2021}. Even with this progress, the best reported linewidth to date is $4\Gamma_\mathrm{TL}$, which required a combination of nanophotonic emission enhancement, electric gating, and resonant driving \cite{wells_coherent_2023}. Reported linewidths using above band excitation in bulk are typically $>50\Gamma_\mathrm{TL}$ (see Supporting Information).

In this paper, we report on a modified SK growth scheme that yields high-quality, telecom-wavelength QDs in InP, using chemical beam epitaxy (CBE). Our approach employs a gallium phosphide (GaP) interlayer to simultaneously decouple the wetting and substrate layers via suppression of uncontrolled interdiffusion, while ensuring ultra-pure growth at densities ranging from 2 to 22 QDs per $\mu$m$^2$. This results in reproducible emission spectra where different excitonic complexes are identifiable, $g^2\left(0\right) = 0.012\pm0.007$ and an upper-bound to the low-power linewidths of only $12.1\pm 6.7\,\Gamma_\mathrm{TL}$ and, in the best case $2.8\pm 1.8\,\Gamma_\mathrm{TL}$. Altogether, these results represent a significant step towards the development of quantum emitters suitable for scalable quantum devices and long-distance communications.

\section{Results and Discussion}


\subsection{Quantum dot growth}

\begin{figure*}[t]
    \centering
    \begin{minipage}{\textwidth}
        \centering
        \includegraphics[width=\textwidth]{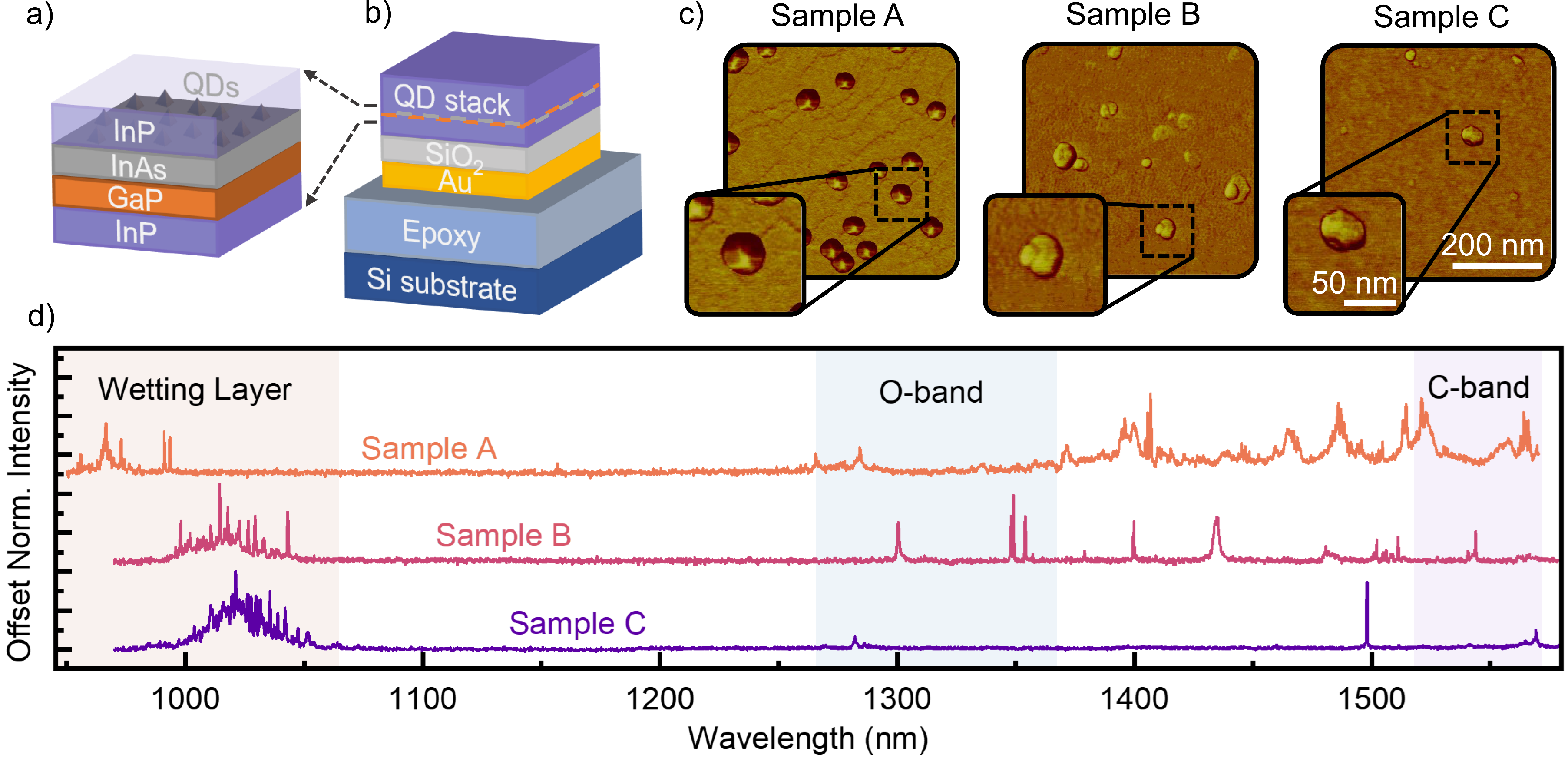}
    \end{minipage}
    \caption{\justifying Telecom-wavelength QDs grown in InP. A diagram of the QD heterostructure used in this work with (a) a Zoom-in of the quantum dot region (highlighting the InAs QDs nucleated atop a GaP interlayer grown on an InP substrate) and (b) the full QD stack, now coated with silica and gold that is then flip-bonded onto a silicon substrate. (c) Atomic force microscopy micrographs of uncapped surface QDs from samples A-C, with insets highlighting a single QD from each sample. (d) Low temperature (4\,K) photoluminescence spectra from each sample grown. The wetting layer emission near 1\,um, along with QD emission in the telecom O- and C-bands are highlighted in the shaded regions.}
    \label{Fig_1}
\end{figure*}

The samples used for these studies were grown using chemical beam epitaxy, with the structure shown schematically in Fig.~\ref{Fig_1}a (see Methods/Supporting Information for details on the growth and processing). In contrast to earlier attempts to grow low-density QDs, here we introduce a 0.4\,nm thick gallium phosphide (GaP) layer between the InP and InAs QD layers, borrowing from growth protocols used to create high-density-QD lasers \cite{poole_growth_2009}. This layer significantly suppresses the exchange of arsenic and phosphorus during the dot growth and ripening phases, effectively decoupling the QDs from the underlying InP layer. This prevented large-scale roughening of the surface and the formation of asymmetric QDs (see Supporting Information Fig. S1). This simple addition now allows us to perform extended growth interruptions (periods of only arsine overpressure to allow for QDs to form) at elevated temperatures, independent of QD size and density. 

Three examples of QDs grown with the GaP interlayer are shown in Fig.~\ref{Fig_1}c. Sample A was grown at a temperature of 545\,°C, with 0.6\,nm of InAs and a 2\,min growth interruption to allow for dot formation and ripening. This yielded very uniform-looking quantum dots with a density of $\sim$22\,$\mu \mathrm{m}^{-2}$. Reducing the growth temperature to 530\,°C and growth interruption to 30\,sec, in sample B, we see a reduced QD density of $\sim$14\,$\mu \mathrm{m}^{-2}$ and the presence of what appear to be incompletely formed quantum dots. Finally, in sample C, with the growth temperature raised back to up 545\,°C, we reduced the thickness of InAs slightly (from 0.6\,nm to 0.5\,nm) and increased the growth interruption to 1\,min. Remarkably, here we measure a QD density of only $\sim$2\,$\mu$m$^{-2}$, consisting of well-formed quantum dots, now absent of any platelet structures. Further details on QD growth (including growth without the GaP interlayer) are provided in the Supporting Information.

Micro-photoluminescence ($\mu$PL) spectra of these samples at cryogenic temperatures (4\,K) are obtained using continuous-wave excitation at $\lambda = 965$\,nm and $\sim$1\,$\mu$m diameter spots. The results, presented in Fig.~\ref{Fig_1}d, confirm that all samples emit in both the telecom O-band (1260\,nm - 1360\,nm) (see Supporting Information) and the C-bands. We also observe wetting layer emission near 1\,$\mu$m. Interestingly, a pronounced blue-shift in the wetting layer emission as the QD density increases indicates that the material from the QDs is drawn from the thinning wetting layer, demonstrating that the suppression of As/P exchange by the GaP interlayer remains effective at elevated temperatures.


\subsection{Quantum dot transitions}
\begin{figure*}[t]
    \centering
    \begin{minipage}{\textwidth}
        \centering
        \includegraphics[width=\textwidth]{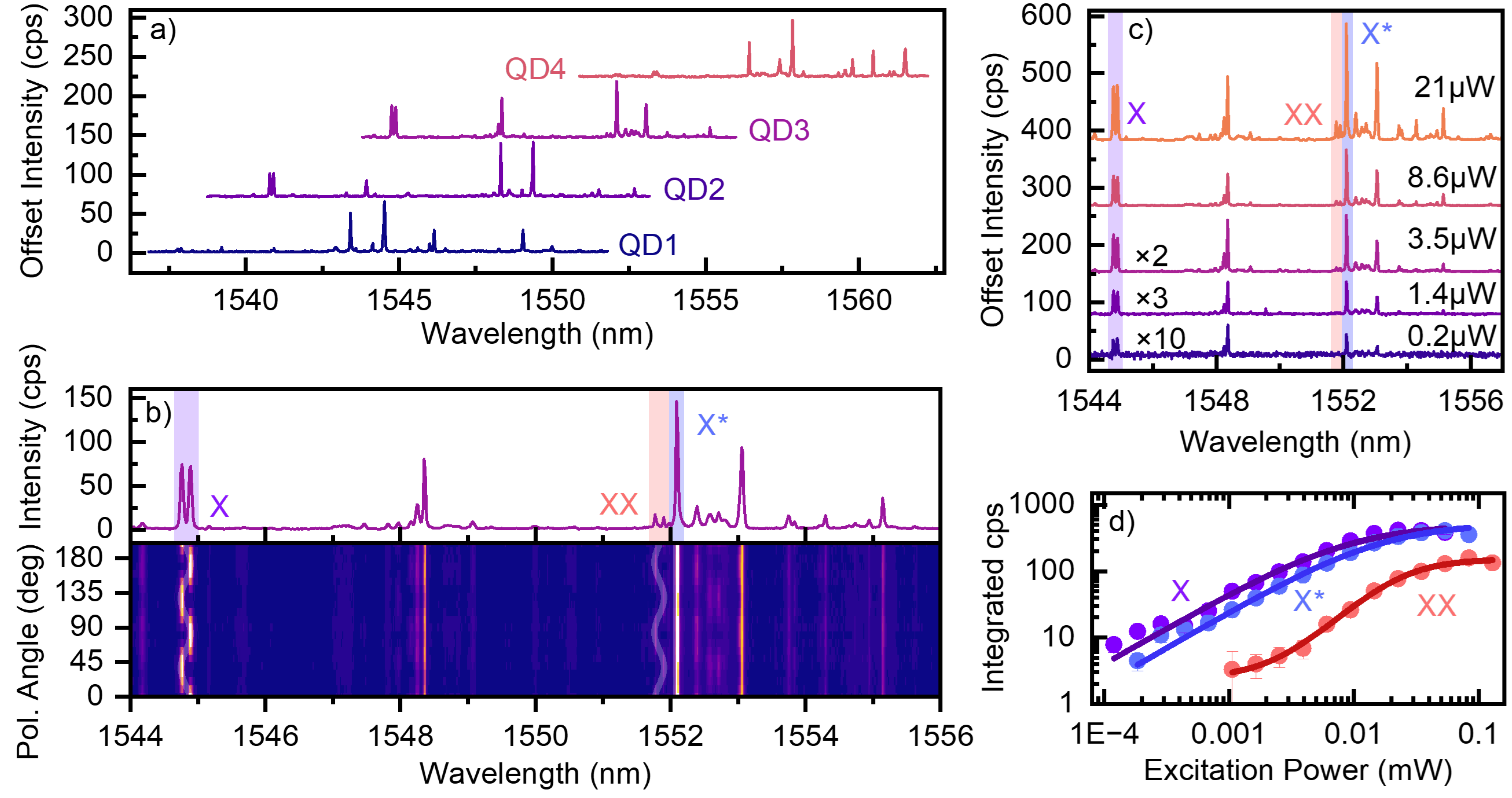}
    \end{minipage}
    \caption{\justifying Single QD spectra. (a) Low temperature $\mu$PL spectra of four randomly selected QDs (labelled QD\,1-QD\,4) emitting in the telecom C-band. (b) Zoom-in on the emission spectrum of QD\,3 with labelled neutral (shaded purple), charged (shaded blue) and bi-exciton (shaded orange) transitions, along with polarization resolved spectral traces with a white overlaid curves that serves as a guide to highlight the orthogonally polarized emission for the three identified complexes. (c) Excitation power-dependent spectra of QD\,3, with the identified transitions again highlighted. (d) Power-dependent emission intensity from each complex, with the measurements given by solid circles and fits by the curves.}
    \label{Fig_2}
\end{figure*}

Next, we study the properties of individual QDs. Since the QDs are emitting from inside a high index material (InP), the collection efficiency is very low, <2\%, making measurements of individual dots very difficult. To overcome this, sample D was grown to allow the fabrication of a weak optical cavity, with the QDs formed under identical conditions to those in sample A. A SiO$_2$/Au mirror was deposited on top of the sample, flipping it over to bond onto a Si carrier wafer, and then removing the substrate (details in Methods)\cite{vajner_-demand_2024, holewa_high-throughput_2024}. This allows us to collect 7-10 times more photons in the wavelength region of interest, providing the count rates necessary for $\mu$PL spectroscopy and time-resolved measurements. Modelling of the weak cavity structure indicated a Purcell modification of <10\%, which is within the dot-to-dot lifetime variation. Spectra from four QDs, taken from sample D, are shown in Fig.~\ref{Fig_2}a. Although each spectrum contains several lines, a close examination reveals almost all can be found in the emission of each QD, identified, for example, by their spacing.

We now focus on a representative QD (QD\,3) to identify the excitonic complexes observed in its spectra (Fig.~\ref{Fig_2}b). To do so, we performed polarization-resolved $\mu$PL measurements (see Methods) to identify the neutral exciton (X) and biexciton (XX). A pair of two closely-spaced transitions is observed with the same energy separation, at 1544.8\,nm and 1551.8\,nm (Fig.~\ref{Fig_2}c), that we attribute to the X and XX, respectively. In both cases, the two lines we associate with the X and XX have orthogonal, linear polarizations. These correspond to the two transition dipoles of each complex, where the separation between the two peaks is the so-called fine structure splitting (FSS). Moreover, we observe that each line of one set is orthogonal to the corresponding line of the second set, a fact that, together with power and lifetime measurements (see below), enables us to confirm their identity.\cite{vajner_-demand_2024} Finally, the brightest peak, found near 1552\,nm shows no polarization dependence, typical of a singly charged trion state ($\mathrm{X^*}$). Based on the residual background doping of our growth system, this is most likely a negatively charged exciton \cite{Reimer_Single_2009}. 

The polarization-resolved spectra presented in Fig.~\ref{Fig_2}b allow us to learn about the asymmetry of the QDs. Although the QDs appear symmetric in the micrographs (Fig.~\ref{Fig_1}c and Supporting Information), a fit of the $\mu$PL traces reveals a FSS of $74\pm 3$\,$\mu$eV. This is lower than typical self-assembled Stranski-Krastanov QDs in this material system, where the FSS is typically >100\,$\mu$eV \cite{birowosuto_fast_2012, holewa_droplet_2022}. We believe, however, that the inherent anisotropy of the dots is likely lower, and the observed value is a result of additional strain introduced during the flip-bonding procedure\cite{lettner_strain-controlled_2021}. Indeed, polarization-resolved measurements on samples grown under identical conditions that have not been flip-bonded reveal a much smaller FSS of $25\pm 4$\,$\mu$eV (see Supporting Information). These smaller values are comparable to recent droplet epitaxy QDs \cite{holewa_droplet_2022, kors_telecom_2018, costa_telecom_2025}.

Further confirmation of peak identification for the excitonic complexes is provided by power-dependent $\mu$PL measurements of QD\,3, shown in  Fig.~\ref{Fig_2}c. In these, we determine how the emission intensity of all three complexes, X, XX and X$^*$, increases with excitation power. This is more evident in Fig.~\ref{Fig_2}d, where we present the integrated counts per second (cps) under each peak as a function of power (in solid circles). We fit this data (curves; see methods for more details), from which we extract the saturation power $\mathrm{P}_\mathrm{sat}$ for each complex and, in the intermediate power region, the rate at which the emission intensity increased in the form of $\mathrm{P}^n$. For QD\,3, for example, both X and $\mathrm{X^*}$ grow linearly, with their emission increasing as $n_\mathrm{X}=1.07\pm0.07$ and $n_\mathrm{X^*}=1.09\pm0.04$, respectively. In contrast, the biexciton emission intensity increases as $n_\mathrm{XX}=1.82 \pm0.15$, growing close to quadratically as expected \cite{sek_applicability_2010}, and confirming that we have correctly identified the complexes (For further confirmation, cross-correlation measurements are provided in the Supporting Information). For this QD, we find $\mathrm{P}_{\mathrm{sat}}^{\mathrm{X}} = 8.9\pm 2.1$\,$\mu$W,  $\mathrm{P}_{\mathrm{sat}}^{\mathrm{X^*}} = 16.4\pm 2.7$\,$\mu$W and $\mathrm{P}_{\mathrm{sat}}^{\mathrm{XX}} = 22.9\pm1.9$\,$\mu$W. These values are typical of the sample, as confirmed by repeated measurements on the different QDs, and in particular for the charged exciton line, which we further explore below, result in a mean saturation power of $\bar{\mathrm{P}}_{\mathrm{sat}}^{\mathrm{X^*}} = 10.6\pm 1.4$\,$\mu$W and $\bar{n} = 1.15 \pm 0.04$.


\subsection{Analysis of single emission lines}
\begin{figure}[htb]
    \begin{center}
    \includegraphics[width=\columnwidth]{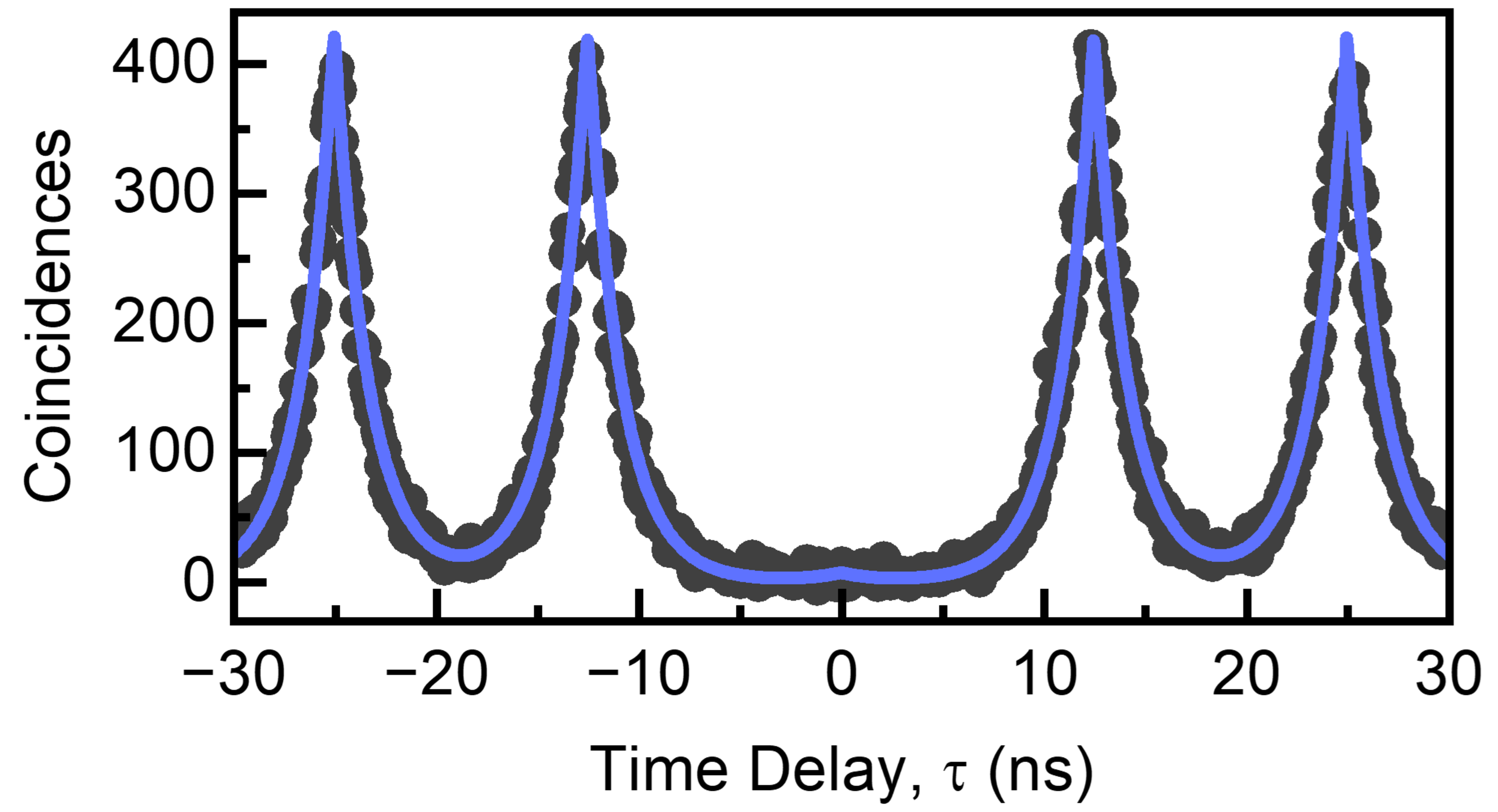}
    \end{center}
    \caption{\justifying Second-order auto-correlation measurement, $g^{(2)}(\tau)$, on the X$^*$ transition from QD\,3. Dark grey circles correspond to measured data points along with a fit (blue line) from which we can extract the emission purity (see Methods).}
    \label{Fig_3}
    
\end{figure}

Having identified the peaks, we now focus on a more detailed study of their emission statistics and linewidths. Ideally, each transition will emit only a single photon each time the dot is excited with a laser pulse. This purity is quantified by the second-order correlation function $g^{(2)}(\tau)$ around $\tau = 0$. We measure this on the X$^*$, which is the only transition sufficiently bright, by exciting into the wetting layer using a tunable 80\,MHz pulsed laser set to $\lambda$ = 965\,nm at an excitation power of approximately $\mathrm{P}_\mathrm{sat}/10$. A fiber-based Hanbury-Brown-Twiss setup is used for the correlation measurements. We plot the measured coincidences in Fig.~\ref{Fig_3}, observing almost no coincidences in the peak centred at $\tau = 0$ relative to the side peaks, indicative of high-purity single-photon emission. For a more quantitative analysis, we fit the data (solid curve; see Methods) and take the ratio of the integrated area from the zero-delay peak relative to the side peaks. In doing so we obtain a raw value of $g^{(2)}(0) = 0.012\pm\mathrm{0.007}$ (See Supporting Information for values used in this calculation).

\begin{figure}
\begin{center}
\includegraphics[width=\columnwidth]{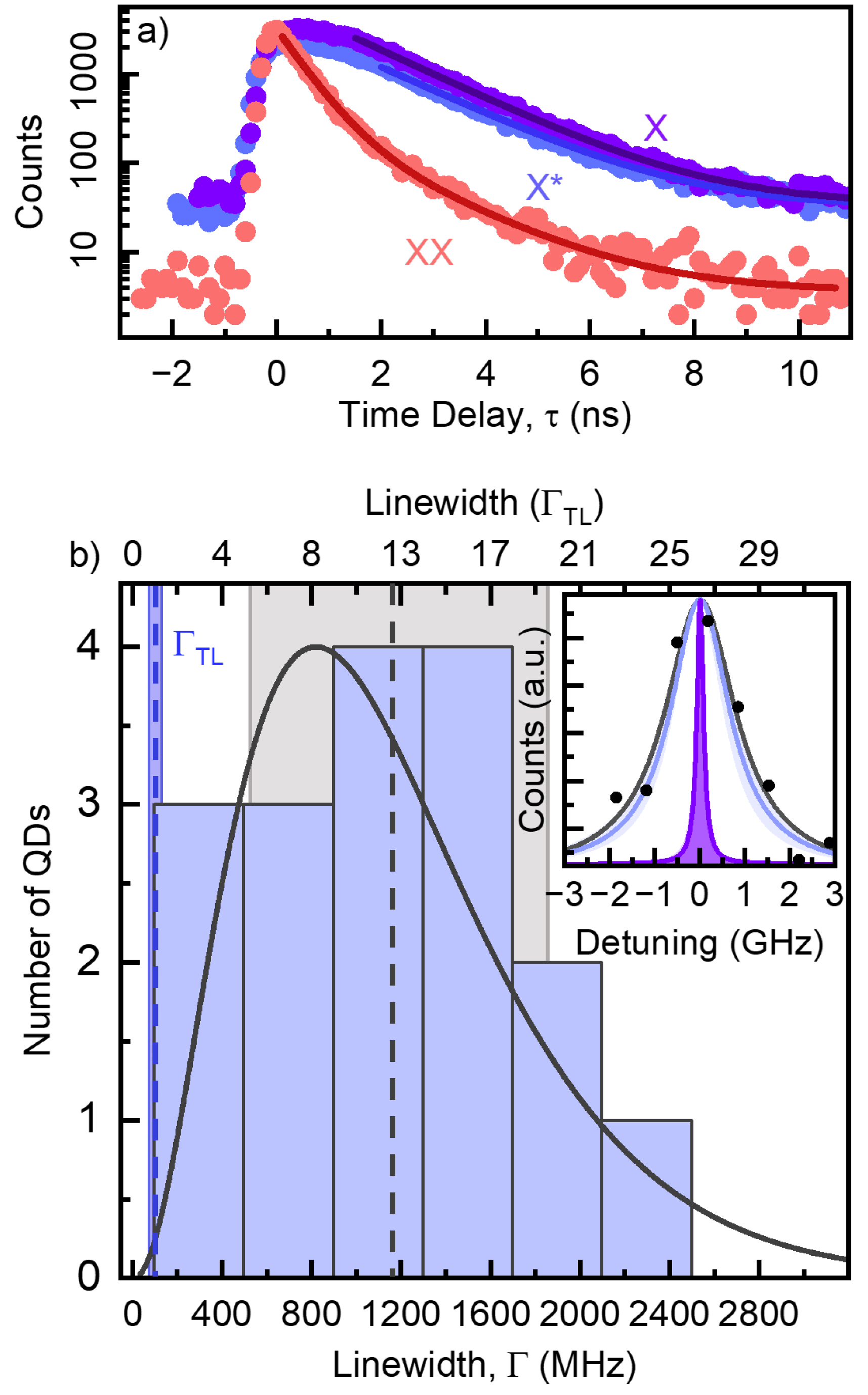}
\end{center}
\caption{\justifying QD lifetimes and linewidths (a) Time-resolved photoluminescence of X, XX and X$^*$ belonging to QD\,3. Data is given by the solid circles and the corresponding fits by the curves. (b) Histogram of extrapolated X$^*$ linewidths from 18 randomly selected QDs. The blue dashed vertical line and shaded region marks $\Gamma_{\mathrm{TL}} = 96\pm25$\,MHz (corresponding to an average lifetime of $\bar\tau_{\mathrm{X^*}} = 1.64\pm0.02$\,ns). The grey dashed line and light grey shaded region show the mean extrapolated linewidth and standard deviation of 1175\,MHz and 648\,MHz, respectively. The top right inset displays the measured linewidth for X$^*$ from our representative QD. Dark circles correspond to measured data points along with Lorentzian fits for measured (dark grey line) and extrapolated linewidths (blue curve) as well as the etalon instrument response function (shaded purple curve).}
\label{Fig_4}
\end{figure}

Having established the single photon purity of the QDs, we now consider their coherence. The transition of a perfectly coherent QD would have a linewidth $\Gamma_\mathrm{TL}$ that is solely determined by its lifetime $T_1$ (as discussed in the introduction). We therefore begin by measuring the time-dependent emission for QDs excited by the pulsed laser, showing typical traces for all three complexes of QD\,3 in Fig.~\ref{Fig_4}a. Here, the X and $\mathrm{X^*}$ decay curves are well fitted by mono-exponential decays, resulting in extracted lifetimes $\tau_\mathrm{X} = 1.59 \pm 0.04$\,ns and $\tau_{\mathrm{X}^*} = 1.57\pm0.04$\,ns. In contrast, the XX decay requires a bi-exponential fit, with a short lifetime corresponding to the XX decay and a faint, long-lived signal that we attribute to a background that arises due to the high power needed to excite the biexciton. This results in a fit of $\tau_\mathrm{XX} = 0.54 \pm 0.24$\,ns, faster than the typical $2\tau_\mathrm{X}$ yet consistent with previous reports on telecom-wavelength QDs \cite{vajner_-demand_2024}. It should be noted that these decay times are expected to be an upper limit since we do not take into account carrier capture and relaxation times in the fitting.

To measure the linewidth of our QDs, we use a free-space scanning etalon (with a resolution of $175\pm25$\,MHz). The results for 17 different QDs are presented in Fig.~\ref{Fig_4}b. In all cases, we measure the $\mathrm{X}^*$, since this is the brightest transition. A typical spectrum (from QD\,3) is shown in the inset. Here, a fit to this raw data results in measured linewidth of $1.8\pm0.4$\,GHz, which we then correct for both the instrument response function (shaded purple curve) and power broadening due to the high excitation power needed (here, at 1.2$\times$\,$\mathrm{P}_{\mathrm{sat}}$) (see Supporting Information for further details); for this QD, we thus obtain a low power linewidth of $\Gamma_{\mathrm{X}^*} = 1597 \pm 279$\,MHz, corresponding to $15.7\pm 2.7\,\Gamma_{\mathrm{TL}}$. We note that this value is an upper bound to the linewidth of the emitter, as our model does not account for the excitation of additional phonons due to the incoherent pumping.\cite{Yao_NonlinearPhotoluminescence_2010} 

We summarize these $\Gamma_{\mathrm{X}^*}$ measurements in the histogram in Fig.~\ref{Fig_4}b, where we also mark $\Gamma_{\mathrm{TL}} = 96\pm25$\,MHz (corresponding to an average lifetime of $\bar\tau_{\mathrm{X^*}} = 1.64\pm0.02$\,ns) with the vertical blue dashed line. Fitting a Gamma distribution to our data (dark grey curve) yields a mean extrapolated linewidth of $\bar\Gamma_{\mathrm{X}^*} = 1175 \pm 648$\,MHz ($12.1\pm 6.7 \, \Gamma_{\mathrm{TL}}$), shown with the dashed grey line and grey shaded region, respectively. Encouragingly, this means that many of our QD transition linewidths are approaching single multiple values of $\Gamma_{\mathrm{TL}}$. In fact, in our random sample of QDs, we find 5 QDs that are within a factor of $10\,\Gamma_{\mathrm{TL}}$, and 2 that are within a factor of 3 of $\Gamma_{\mathrm{TL}}$, with a narrowest linewidth of only $272\pm 171$\, MHz $\left(2.8\pm 1.8 \, \Gamma_{\mathrm{TL}} \right)$.

\subsection{Conclusion}
In summary, we have introduced a modified Stranski-Krastanov growth protocol to create high-quality telecom-wavelength InAs QDs in InP. By decoupling the substrate and QD layers through the addition of a GaP interlayer, we demonstrate growth of self-assembled QDs at extremely low densities that have physical and optical properties more typical of DE QDs rather than previous SK ones. For example, we observe highly symmetric QDs, with aspect ratios of 0.86 (see Supporting Information), comparable to DE values $\left(\approx0.8-0.9\right)$ and well above $\approx0.5$ that is typical of SK dots \cite{skiba-szymanska_universal_2017, holewa_droplet_2022}. We note that the InP capping step is known to introduce anisotropic effects due to differing surface diffusion rates along the different crystallographic axes \cite{poole_chemical_2001}, meaning that with optimization of the capping step, we expect further improvement of the already small fine structure splitting of our QDs.

Notably, using the relatively high powers needed to get sufficient counts (1-6$\times\mathrm{P_{sat}}$), we are able to measure the transition linewidths of our QDs in the telecommunication C-band. However, since the quality of emitters is revealed at the low-power excitation regime, where there is no excitation-induced power broadening, we account for this effect using a rigorous thermal bath model\cite{Yao_NonlinearPhotoluminescence_2010} that, nevertheless does not include pump-power dependent phonon excitation\cite{Laferriere_Approaching-transform_2023}. Neglecting this additional phonon-induced broadening, we extract an upper-bound to the low-power QD transition linewidths, finding that they begin to approach the transform limit for ungated QDs in bulk InP, even with this above-band excitation. In fact, we find a low-power statistical mean linewidth of only $12.1\pm 6.7 \, \Gamma_{\mathrm{TL}}$, and a narrowest transition of only $2.8\pm 1.8 \, \Gamma_{\mathrm{TL}}$. This is in stark contrast to other approaches that, in the same conditions, report on linewidths in the $50 - 100 \, \Gamma_{\mathrm{TL}}$ range \cite{musial_high-purity_2020, phillips_purcell-enhanced_2024, rahaman_efficient_2024}.

Several techniques have been developed to improve QD transition linewidths, all of which are compatible with our emitters. First, the QD emission rate may be enhanced using photonic resonators, such as photonic crystal cavities \cite{phillips_purcell-enhanced_2024} or circular Bragg gratings \cite{kaupp_purcell-enhanced_2023, barbiero_high-performance_2022}, effectively broadening their transitions so that they are less affected by noise. Second, resonant excitation schemes directly populate the desired QD levels \cite{vajner_-demand_2024, nguyen_ultra-coherent_2011, muller_resonance_2007}, reducing the injection of charges into the solid-state matrix around the emitter and hence leading to the observation of narrower linewidths. Similarly, electronic gating is known to stabilize the charge environment \cite{pedersen_near_2020, wells_coherent_2023, martin_stark_2025}, dramatically reducing linewidths. We summarize the results of these techniques on telecom-wavelength QDs in InP in the Supporting Information, noting here that by applying \textit{all} of these simultaneously researchers have reduced linewidths from $\approx50\,\Gamma_{\mathrm{TL}}$ \cite{skiba-szymanska_universal_2017} to $\approx4\,\Gamma_{\mathrm{TL}}$ \cite{wells_coherent_2023}. Despite this dramatic 13-fold reduction in linewidth, we observe narrower low-power linewidths in $12\%$ of our ungated (c.f. Fig.~\ref{Fig_4}b), bulk QDs with above band excitation. This suggests that our approach may well lead to the transform-limited C-band transitions required for coherent quantum photonics or the deterministic generation of indistinguishable photons needed by emerging quantum technologies.

\section{Methods}

\textbf{QD Sample Growth}:
The samples were grown on semi-insulating InP:Fe (001) substrates by chemical beam epitaxy (CBE) using trimethyl-indium and pre-cracked phosphine and arsine for the indium, phosphorus, and arsenic sources, respectively. The substrate temperature was controlled using band-edge thermometry \cite{roth_Realtime-Control_1999}. Although the absolute calibration may not be accurate, the temperature was highly reproducible, a requirement for good dot growth control. The native oxide was removed at 580\,°C followed by the growth of a 200 nm InP buffer at 530\,°C and a short ramp to the dot growth temperature. Dot growth was initiated by either directly depositing InAs (at a growth rate of 0.1\,nm/s) on the InP or first depositing a nominal 0.4\,nm of GaP followed by the InAs. A growth interruption (i.e. no indium was being supplied to the chamber) with an As overpressure allowed migration of indium on the surface to form the dots and those dots to ripen.

The dots were then capped with 20\,nm of InP before the temperature was ramped back to 530\,°C for a 90\,nm InP spacer. The dot growth was repeated as above with the substrate heater being turned off at the end of the growth interruption and the sample cooled down under an As overpressure.

\textbf{Flip-Bonding Procedure}:
To enhance the collection efficiency of the QD emission, sample D was grown (with the same growth conditions as Sample A) to be flip-bonded onto a mirror to create a weak cavity. This increased the light collection by a factor of 7 to 10 times in the 1.5\,$\mu$m wavelength range with a minimal Purcell modification of <10\%. A sacrificial InGaAs layer 200\,nm thick was first grown before 472\,nm of InP which contained a single layer of InAs QDs in the middle. This wafer was then coated with 250\,nm of SiO$_2$ followed by 250\,nm of Au. Pieces were then cleaved from this wafer and bonded Au side down onto a Si wafer using EPO-TEK\textsuperscript{\textregistered} 301 two-part optical epoxy. The InP substrate was removed using HCl and the InGaAs then selectively etched away using a dilute H$_2$SO$_4$ and H$_2$O$_2$ aqueous solution. 

\textbf{Low Temperature Polarization Resolved Photoluminescence Spectroscopy}: 
Spectrally resolved photoluminescence measurements were performed at 4\,K using a closed-cycle helium cryostat. Quantum dots were excited via the wetting layer using continuous-wave excitation at a wavelength of $\lambda = 965$\,nm through a 100x cryogenic objective (numerical aperture = 0.81, spot size of $\approx$1\,um). The emitted PL was collected through the same objective, coupled into an SMF-28 fiber, and directed to a spectrometer equipped with a liquid nitrogen-cooled InGaAs linear array detector.

For polarization-resolved measurements, a rotating half-wave plate followed by a fixed linear polarizer was placed in the free-space collection path prior to fiber coupling. The polarization dependence of the QD emission was analyzed by rotating the half-wave plate while keeping the linear polarizer fixed. Emission peaks of interest were fitted with Lorentzian functions to extract their center wavelengths, which were subsequently fitted to a sine function to determine the amplitude (peak-to-peak) of the fine-structure splitting.

\textbf{Power-Dependent $\mu$PL Measurements}:
QD emission spectra were fitted with Lorentzian functions at each excitation power. The intensity $I(\mathrm{P})$ for a given emission peak was calculated by multiplying the fitted peak area by the spectrometer resolution (0.02\,nm), converting it to total counts. To determine the saturation power $\mathrm{P}_\mathrm{sat}$ and power dependence exponent $n$, the resulting intensity values were fitted using the following saturation model\cite{stufler_Power_2004}:

\begin{align}
    I(\mathrm{P}) = I_\mathrm{sat} \cdot \frac{\mathrm{P}^n}{\mathrm{P}_\mathrm{sat}^n + \mathrm{P}^n} + C
\end{align} where $I_\mathrm{sat}$ is the maximum emission intensity at saturation, and $C$ accounts for a constant background offset.

\textbf{Single-Photon Purity Analysis}: QDs of interest are excited by pumping into the wetting layer using a tunable 80 MHz pulsed laser set to $\lambda$ = 965 nm, operated at approximately $\mathrm{P}_\mathrm{sat}$/10. QD emission was then coupled into the collection fiber and spectrally filtered using a 0.1\,nm tunable band pass filter to isolate individual emission lines. The filtered signal was then directed to a fiber-based Hanbury-Brown-Twiss setup, where it was split by a 50:50 fiber-coupled beam splitter and detected by two superconducting nanowire single-photon detectors. Coincidence events between the two detectors were recorded as a function of the time delay $\tau$. To analytically quantify the single-photon purity of these sources, the measured coincidence histogram was fitted using the following equation\cite{holewa_high-throughput_2024},

 \begin{align}
 g^{(2)}(\tau) &= C_0 \left[ 
     e^{- \frac{|\tau|}{T_1}} \right] + C_1 \sum_{n \ne 0} \left[e^{- \frac{|\tau - n \tau_{\mathrm{rep}}|}{T_1}}  
     \right] + C_b
 \end{align} where $C_0$ is the amplitude of the zero-delay peak due to multi-photon emission events, $T_1$ is the radiative decay time, $C_1$ corresponds to the amplitude of the side peaks at time delays equal to $\tau=n\tau_\mathrm{rep}$ for n$\ne$0, and finally $C_\mathrm{b}$ is a constant background level. To determine the single-photon purity, we compare the integrated area of the zero-delay peak to that of the adjacent side peaks (fitted parameter values are provided in the Supporting Information).\\

\textbf{SUPPORTING INFORMATION}
Additional details on the growth of QD samples, optical characterization and fine-structure splitting measurements on non-flip-bonded QDs (emitting in both the telecom O- and C-bands), further explanation on linewidth power-broadening, along with a summary of current reports on telecom emitting InAs/InP QD linewidths. (DOC)\\

\textbf{ACKNOWLEDGEMENTS}
This work was supported by the National Research Council of Canada through the Quantum Sensing Challenge Program project `Telecom Photonic Resources for Quantum Sensing' (QSP-061). The authors gratefully acknowledge the support from the Canadian Foundation for Innovation, the National Science and Engineering Research Council and Queen's University. Finally, the authors would like to thank Stephen Hughes for informative conversations regarding linewidth broadening under incoherent excitation.\\

\textbf{DATA AVAILABILITY}
The data that support the findings of this study are available from the corresponding author upon reasonable request.

\onecolumn
\bibliography{main}

\end{document}

\pagebreak

In this supplemental material we provide additional details on the growth of several quantum dot (QD) samples with and without a gallium phosphide layer, time resolved photoluminescence measurements on QDs emitting in both the telecom O-band and C-band from QD sample A, additional second-order correlation measurements from the representative QD in text to further confirm correct excitonic assignment, and finally a current summary on linewidth measurements reported on InAs/InP QDs. 

\subsection{Additional details on QD growth and flip-bonding procedure}

\begin{table*}[htb]
    \centering
    \caption{Growth conditions for planar QD samples (listed in order of sample creation), including the presence of a GaP interlayer, temperature, InAs amount, and growth interruption (GI) time, along with measured QD density, mean diameter and aspect ratio.}
    \begin{tabular}{cccccccc}
        \toprule
        \textbf{Sample ID} & \textbf{GaP} & \textbf{Temp.} & \textbf{InAs Amount} & \textbf{GI Time} & \textbf{Density} & \textbf{Mean Dia.} & $\mathbf{R_x/R_y}$ \\
        \midrule
        S & no  & 530\,°C & 0.6\,nm & 30 sec & 40\,$\mu$m$^{-2}$ & 62\,$\pm$\,30\,nm & 0.54\\
        B & Yes & 530\,°C & 0.6\,nm & 30 sec & 14\,$\mu$m$^{-2}$ & 39\,$\pm$\,20\,nm & 0.71\\
        A & Yes & 545\,°C & 0.6\,nm & 2 min  & 22\,$\mu$m$^{-2}$ & 62\,$\pm$\,20\,nm & 0.86\\
        C & Yes & 545\,°C & 0.5\,nm & 1 min  & 2\,$\mu$m$^{-2}$  & 48\,$\pm$\,32\,nm & 0.70\\
        \bottomrule
    \end{tabular}
    \label{growthtable}
\end{table*}

\begin{figure*}[htb]
    \centering
    \begin{minipage}{\textwidth}
        \centering
        \includegraphics[width=\textwidth]{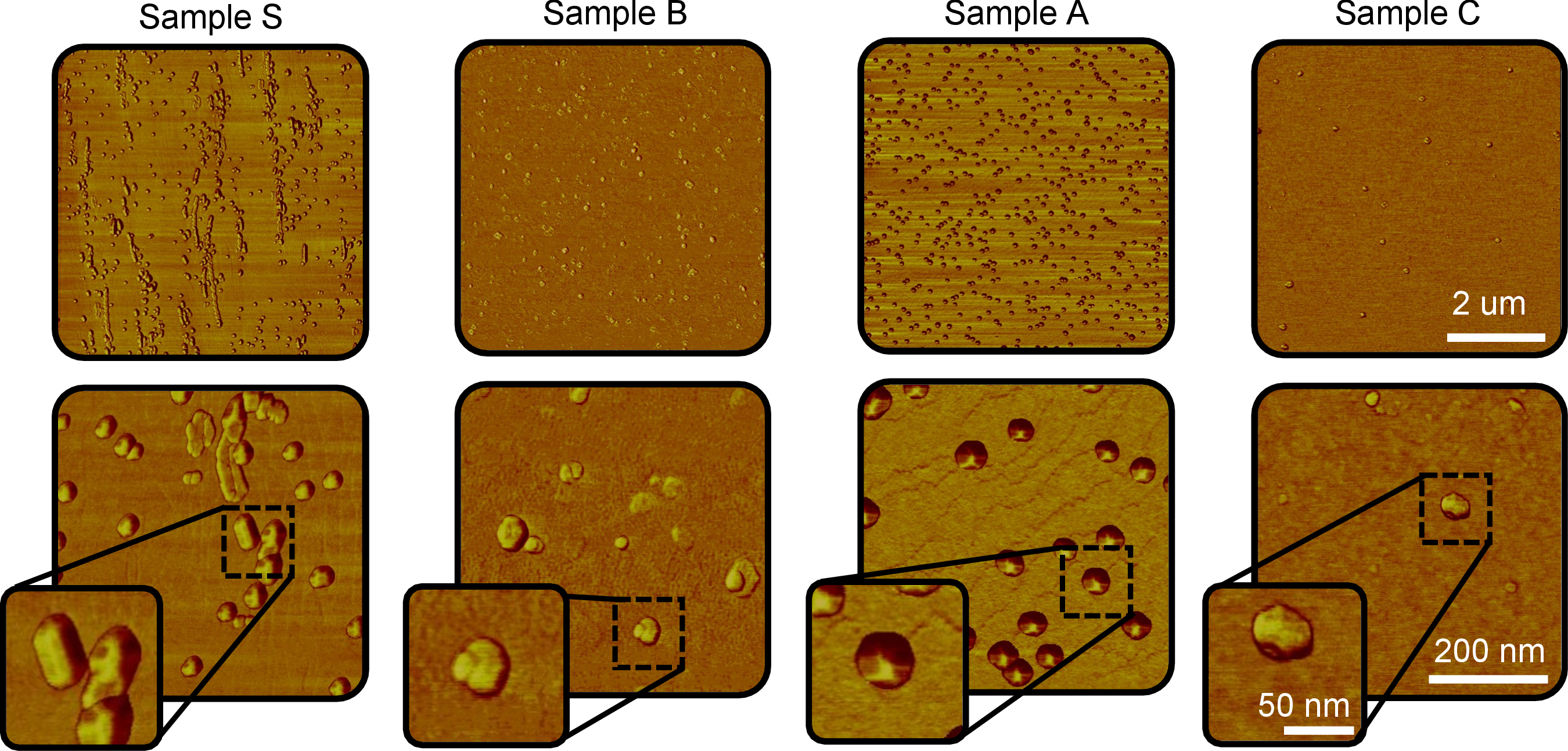}
    \end{minipage}
    \caption{Telecom-wavelength QDs grown in InP. Atomic force microscopy micrographs of uncapped surface QDs from samples A to D for decreasing length scales (top to bottom). Insets highlight a single QD from each sample.}
    \label{fig:growth}
\end{figure*}

 Across four QD samples labelled S, A, B and C, we systematically varied the growth temperature, precursor gas fluxes, and growth interruption times. Growth parameters are summarized in Table~\ref{growthtable}, along with additional growth details/procedures provided in the Methods. Each sample consists of a layer of uncapped surface QDs grown overtop a buried QD layer, whose morphology directly correlates with the underlying QDs. While not optically active, these surface dots indicate changes in both buried and surface QD morphology under varying growth conditions. Atomic force microscopy (AFM) images presenting surface morphology for each sample are shown in Fig.~\ref{fig:growth}. The AFM system used in this work was a Veeco-Bruker Extended MultiMode Nanoscole IIIa, in Tapping Mode, with TESPA-V2 Al backside-coated, n-doped Si tips. To analyze the physical properties of the quantum dots, like density, size, and aspect ratio, we used the AFM characterization software NanoScope IIIa (Bruker) for each scan. Here, the automated software identifies each QD based on selected height bounds to determine the number of QDs for a given scan area. The software then determines the average diameter, length (longest axis) and width (perpendicular bisector to length) of each dot. Using these values, for each scan, we calculate the mean aspect ratio by taking the ratio of the mean QD length and width.

In baseline growth conditions (sample S), AFM characterization reveals uneven surface morphology exhibiting hillocks and elongated dots. Control experiments omitting the buried QD layer resulted in a smooth surface, indicating that these features are the result of the dot growth process used in sample S. For this sample, the InAs was deposited directly onto the InP surface, which is expected to result in unwanted As/P exchange and roughening. This uneven morphology is then carried forward in subsequently grown layers. While this uneven distribution of surface dots may not reflect the surface morphology of dots below, this suggests a mechanism for additional growth due to the unintended generation of InAsP \cite{poole_growth_2009}.

To suppress this As/P intermixing, in sample B, we introduced a 0.4\,nm thick pseudomorphic GaP interlayer between the InP buffer and InAs wetting layers. AFM characterization confirmed that the introduction of the GaP interlayer was not only able to maintain smooth surface morphology, but also resulted in a reduction in quantum dot density from approximately 40\,$\mu$m$^{-2}$ to 14\,$\mu$m$^{-2}$ for the same amount of deposited InAs. Additionally, the mean diameter of the QDs was also reduced by approximately 37\%. The resultant dots exhibited platelet-like morphologies, consistent with a ripening process that is kinetically limited, requiring a longer growth interruption time to form completely.

\begin{figure*}[t]
    \centering
    \begin{minipage}{\textwidth}
        \centering
        \includegraphics[width=\textwidth]{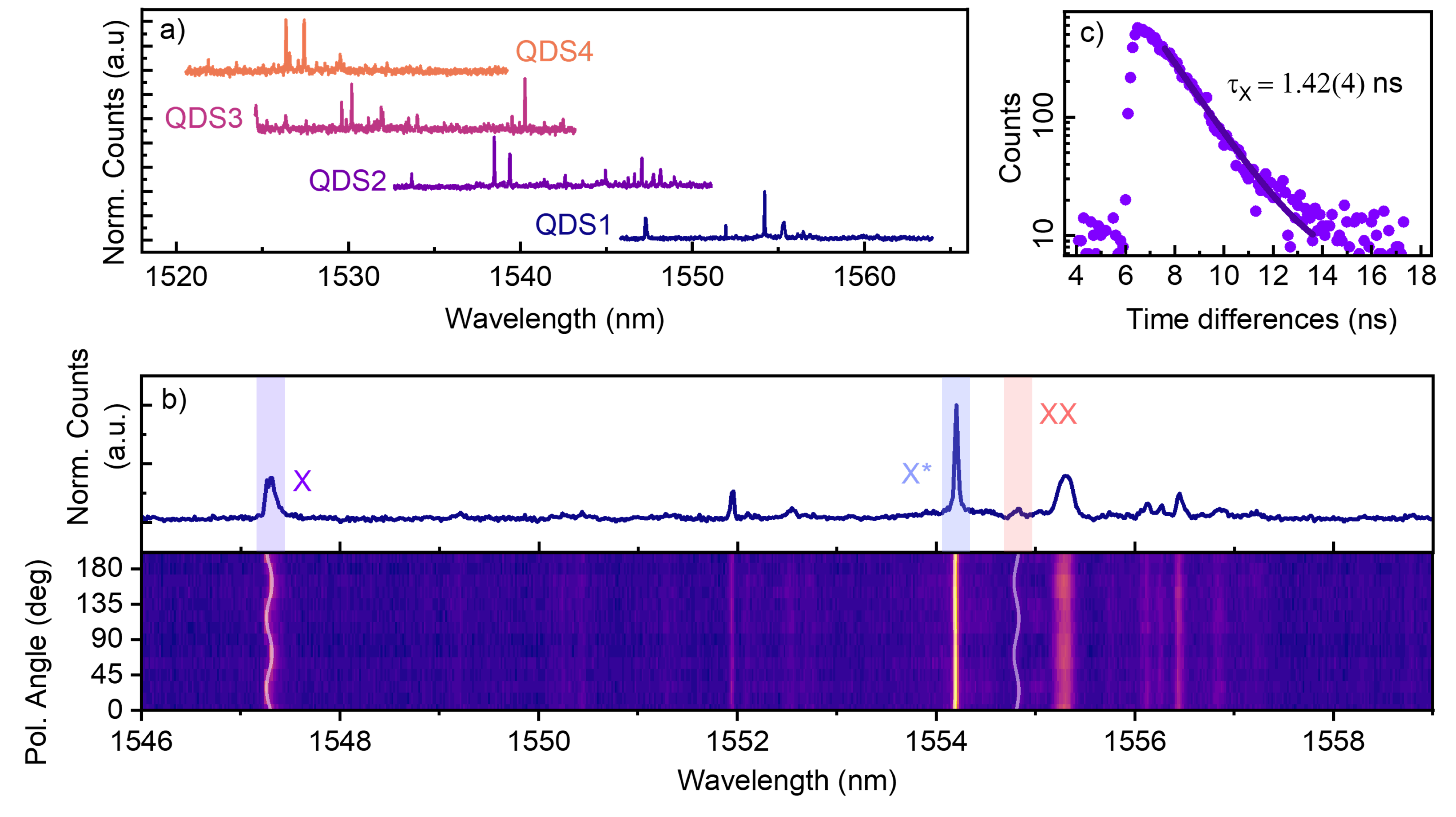}
    \end{minipage}
    \caption{(a) Low temperature $\mu$PL spectra for four randomly selected telecom C-band emitting QDs (labelled S1-S4) that have not undergone the flip-bonding process. (b) Zoom-in spectra of QDS1 with labelled neutral (purple highlight), charged (blue highlight) and bi-exciton (orange highlight), along with a polarization resolved spectral map of QDS1 with fitted sinusoids (overlaid white lines) for the three identified complexes. (c) Time-resolved photoluminescence of X belonging to QDS1 exhibiting a radiative decay time similar to X complexes measured on flip-bonded QDs.}
    \label{fig:Cband_PL}
\end{figure*}

With the GaP interlayer effectively stabilizing surface composition, we explored growth at elevated temperatures (15\,°C above sample B) combined with extended growth interruption periods (increased from 30\,sec to 2\,min) in sample A. This resulted in a controlled increase in dot density, improved dot uniformity (now absent of platelet features) and QD size. In sample C, we aimed to minimize QD density by slightly reducing the InAs deposition amount (from 6\,sec to 5\,sec of injection time) and shortening the growth interruption time to 1 minute. Yielding a QD density as low as 2~$\mu$m$^{-2}$, this demonstrates how by using the GaP interlayer, we can target a wide range of QD densities while maintaining consistent physical QD properties.

\subsection{Additional Optical Measurements on Non-Flip-bonded QDs}

\begin{figure}[]
    \begin{center}
    \includegraphics[width=\columnwidth]{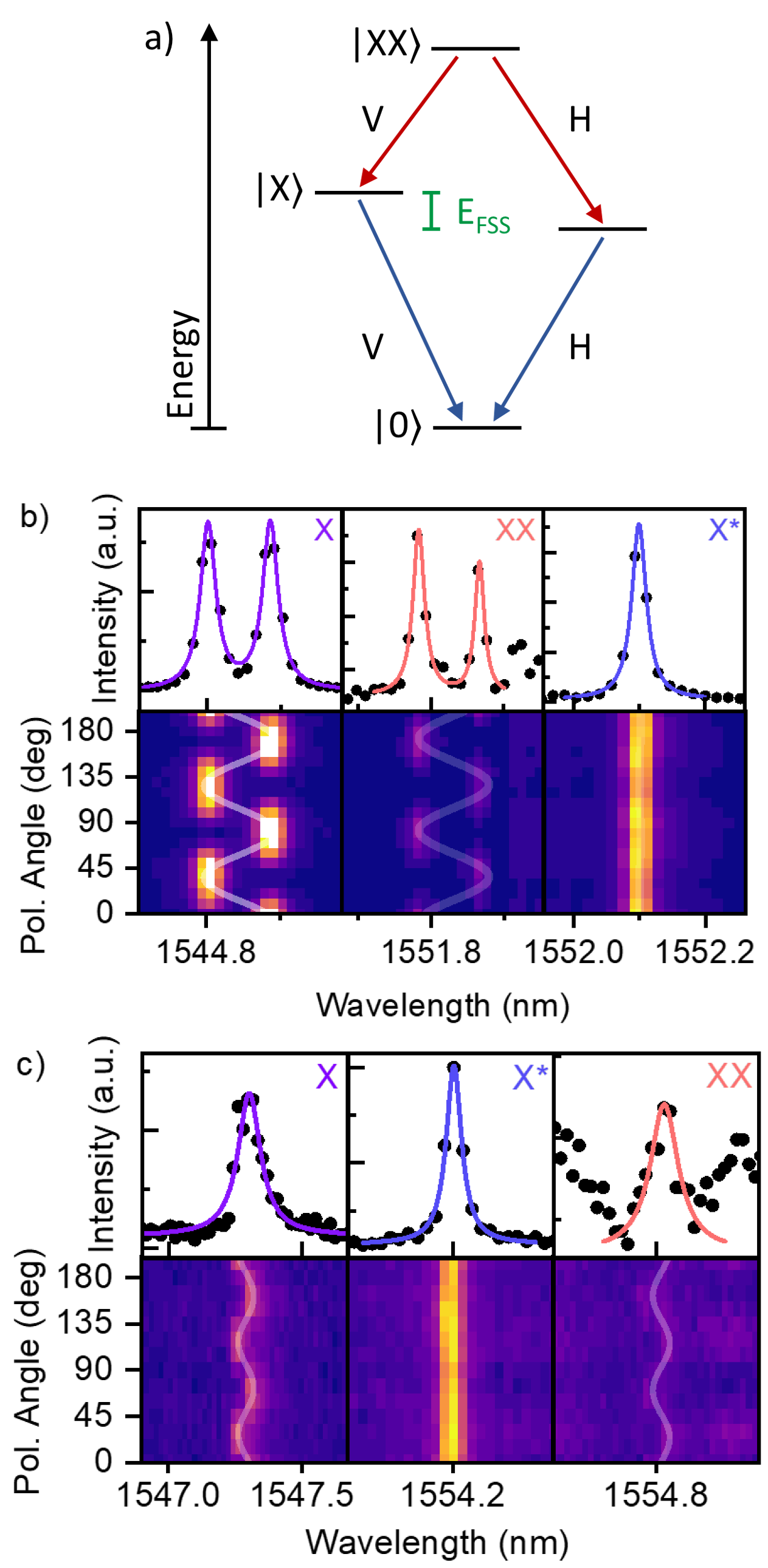}
    \end{center}
    \caption{QD fine structure splitting. (a) Energy level structure for the XX-X radiative cascade in a quantum dot. The bi-exciton recombines to one of two orthogonally polarized exciton states H or V, which are split by a non-zero fine structure splitting (E$_{FSS}$). Each exciton subsequently recombines to the ground state, emitting a second photon with the same polarization as the XX. We highlight the energy difference in H and V polarized photons by showing zoom-in spectra of the X, XX, and X* complexes of (b) QD\,3 and (c) QD\,S1, which exhibit varying degrees of FSS (white overlay fits).}
    \label{fig:FSS}
\end{figure}

\begin{figure*}[t]
    \centering
    \begin{minipage}{\textwidth}
        \centering
        \includegraphics[width=\textwidth]{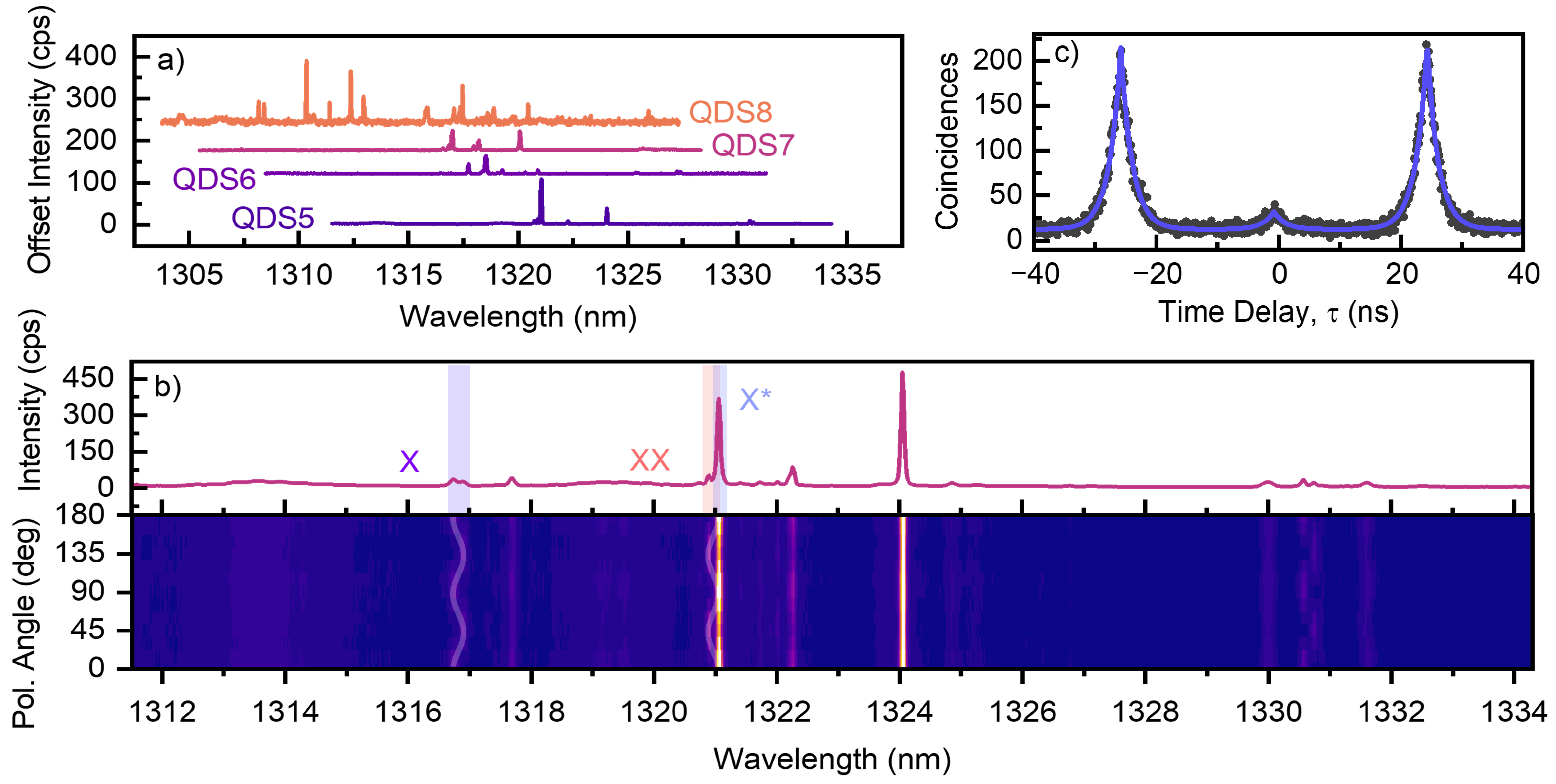}
    \end{minipage}
    \caption{(a) Low temperature $\mu$PL spectra for three randomly selected telecom O-band emitting QDs (labelled S5-S8) that have not undergone the flip-bonding process. (b) Zoom-in spectra of QD\,S5 with labelled neutral (purple highlight), charged (blue highlight) and bi-exciton (orange highlight), along with a polarization resolved spectral map of QD\,S5 with fitted sinusoids (overlaid white lines) for the three identified complexes. (c) Second-order auto-correlation measurement, $g^{(2)}(\tau)$, on the X$^*$ transition at 1310.35\,nm from QD\,S8. Dark grey circles correspond to measured data points along with a fit (blue line) from which we can extract the emission purity (see Methods).}
    \label{fig:Oband_PL}
\end{figure*}

Fig.~\ref{fig:Cband_PL}a shows four randomly selected QDs from a sample A emitting in the telecom C-band. These dots exhibit a similar spectral profile to each other along with those displayed in the main text (c.f. Fig.\,2a). Performing polarization resolved spectroscopy (described in the main text) on QD\,S1, we see again pairs of orthogonally polarized peaks at 1547.26\,nm and 1554.82\,nm, along with a peak that exhibits no polarization dependence at 1554.20\,nm. Based on prior characterization of these QDs, we label these complexes X, XX and X$^*$ respectively. In QD\,S1, using the fitted sinusoid method (shown as white overlays in Fig.~\ref{fig:FSS}c), we extract a FSS of 25$\pm$4\,$\mu$eV. Comparing this smaller FSS value to that of QD\,3 (discussed in the main text and shown in Fig.~\ref{fig:FSS}b) provides evidence that additional anisotropic strain is imparted onto the QD sample during the flip-bonding procedure. Performing time-resolved PL measurements on X, we extract a radiative decay time of 1.42$\pm$0.21\,ns from mono-exponential fitting. Comparing this value to a decay time measured on flip-bonded QDs, we can likely conclude that minimal, if any, lifetime enhancements due to weak cavity effects are observed, as expected from modelling.

\subsection{Fine Structure Splitting Analysis}

As shown in Fig.~\ref{fig:FSS}a, there are two decay pathways from the bi-exciton state to the ground state, which will produce orthogonally polarized photons aligned to one of the two in-plane emission dipoles of the QD. For a perfectly symmetric quantum dot, these two decay pathways are indistinguishable from one another and result in degenerate H- and V-polarized XX and X photons. It is very often the case that in real quantum dots, a non-zero fine structure splitting arises from an anisotropic confinement potential of charge carriers as a result of either an asymmetric QD shape or strain imparted onto the QD.\cite{laferriere_Systematic_2021} This breaks the rotational symmetry of the QD, lifting the degeneracy in emission energy for orthogonally polarized photons. It is for this reason that we believe the flip-bonding procedure may have increased the FSS of our sources by mechanically straining the dots. In Fig.~\ref{fig:FSS}b and c, we show two zoom-in spectra on the X, X* and XX emission lines from QD\,3 and QD\,S1, respectively. In QD\,3, both orthogonally polarized emission peaks in the X and XX complexes can be clearly resolved and exhibit an out-of-phase relationship. This is in contrast to the X and XX complexes of QD\,S1. Rather than pairs of distinct peaks, here we see single emission lines whose center wavelength appears to vary as a function of polarization angle (selecting for each orthogonal emission line). 

To determine the FSS from the polarization-resolved spectroscopy, two methods were used. In the first method, when both orthogonal emission peaks could not be resolved (as is the case for QD\,S1), a single Lorentzian peak was fit to the emission spectra for each polarization angle. Plotting the center wavelength vs polarization angle and fitting this to a sinusoid, the FSS was then determined via the peak-to-peak amplitude from the fit. When both orthogonal emission peaks \textit{could} be well resolved on the spectrometer, a double Lorentzian peak function was instead used to fit both peaks for each polarization angle. While a sinusoid could be used to again determine the FSS (which we use to guide the eye in polarization-resolved heat-maps, as in Fig.~\ref{fig:FSS}b), in these cases, the FSS was instead directly given by the spectral distance between each pair of peaks.

\subsection{O-Band Telecom Quantum Dots}
We also performed optical measurements on non-flip-bonded quantum dots emitting in the telecom O-band. We found that QDs emitting in this wavelength range were significantly brighter than their C-band counterparts. We believe this is partly due to the greater spectrometer grating efficiency for this wavelength range, and also potentially due to a lack of additional higher charge states (as can be seen in the cleaner emission spectra of these QDs in Fig.~\ref{fig:Oband_PL}a), meaning emission intensity is distributed across fewer excitonic states. Again, from polarization resolved spectroscopy on QD\,S7 we tentatively identify the neutral (1316.82\,nm), charged (1321.06\,nm) and bi-exciton (1320.90\,nm) complexes based on prior QD characterization. Interestingly, it appears that the XX and X* emission lines partially overlap (yet can be distinguished based on their relative intensity as a function of polarization angle), and it was not uncommon to see this in other O-band emitting dots. Fitting the orthogonally polarized traces, we extract a FSS of 13.9$\pm$0.5\,$\mu$eV (white overlays in Fig~\ref{fig:Oband_PL}b).

On QD\,S8 we measured the single-photon purity of the charged exciton line close to 1310\,nm (c.f. Fig.~\ref{fig:Oband_PL}c) by exciting above-band with a 670\,nm pulsed laser operating at close to P$_{\mathrm{sat}}$ at a rep rate of 25\,ns (see Methods for further details). Fitting the coincidences to Eq.\,2 in the main text (fit values are provided in Table~\ref{G2_fit_table}), a single-photon purity around the zero time delay was calculated to be $g^{(2)}(0) = $ 0.210$\pm$0.011, which is reduced to $g^{(2)}(0) = 0.096\pm\mathrm{0.007}$ after background subtraction.

\begin{table*}[htb]
    \centering
    \caption{Fitted parameters used in single-photon purity calculations (parameter definitions can be found in the Methods)}
    \begin{NiceTabular}{ccc}
        \toprule
        \textbf{Fit Parameter} & \textbf{O-Band} & \textbf{C-Band} \\
        \midrule
        C$_0$ & 41.5$\pm$2.1 & 5.5$\pm$3.1 \\
        C$_0$\tabularnote{Background corrected} & 19.7$\pm$1.5 & 5.5$\pm$3.1 \\
        C$_1$ & 197.4$\pm$2.0 & 428.1$\pm$2.4 \\
        C$_1$\tabularnote{Background corrected} & 203.3$\pm$1.4 & 428.4$\pm$2.4\\
        C$_b$ & 12.4$\pm$0.3 & 1.1$\pm$1.1 \\

        \bottomrule
    \end{NiceTabular}
    \label{G2_fit_table}
\end{table*}

\subsection{Additional Second-order Correlation Measurements on QD\,3}
To correctly identify which emission lines belong to specific QD transitions, we performed a series of polarization-resolved and power-dependent $\mu$PL measurements, along with time-resolved photoluminescence (TRPL) measurements. While together these results provide us with good confidence that we've correctly identified the appropriate complexes, both the power-dependent and TRPL measurements revealed similar excitation-dependent intensities and decay times. To further distinguish between the excitonic complexes, we measured second-order correlations between the tentatively assigned X, X* and XX complexes from QD\,3. To do this, each of the spectral lines was filtered out using a 0.1\,nm band-pass filter and directed to separate SNSPDs. Coincidence detection measurements were then recorded between photons from either X-X*, X-XX, and X*-X*, with the trigger starting on either the XX or X*, and stopping on X (in the case of cross-correlation measurements), or X* (for the auto-correlation measurement). 

\begin{figure}[htb]
    \begin{center}
    \includegraphics[width=\columnwidth]{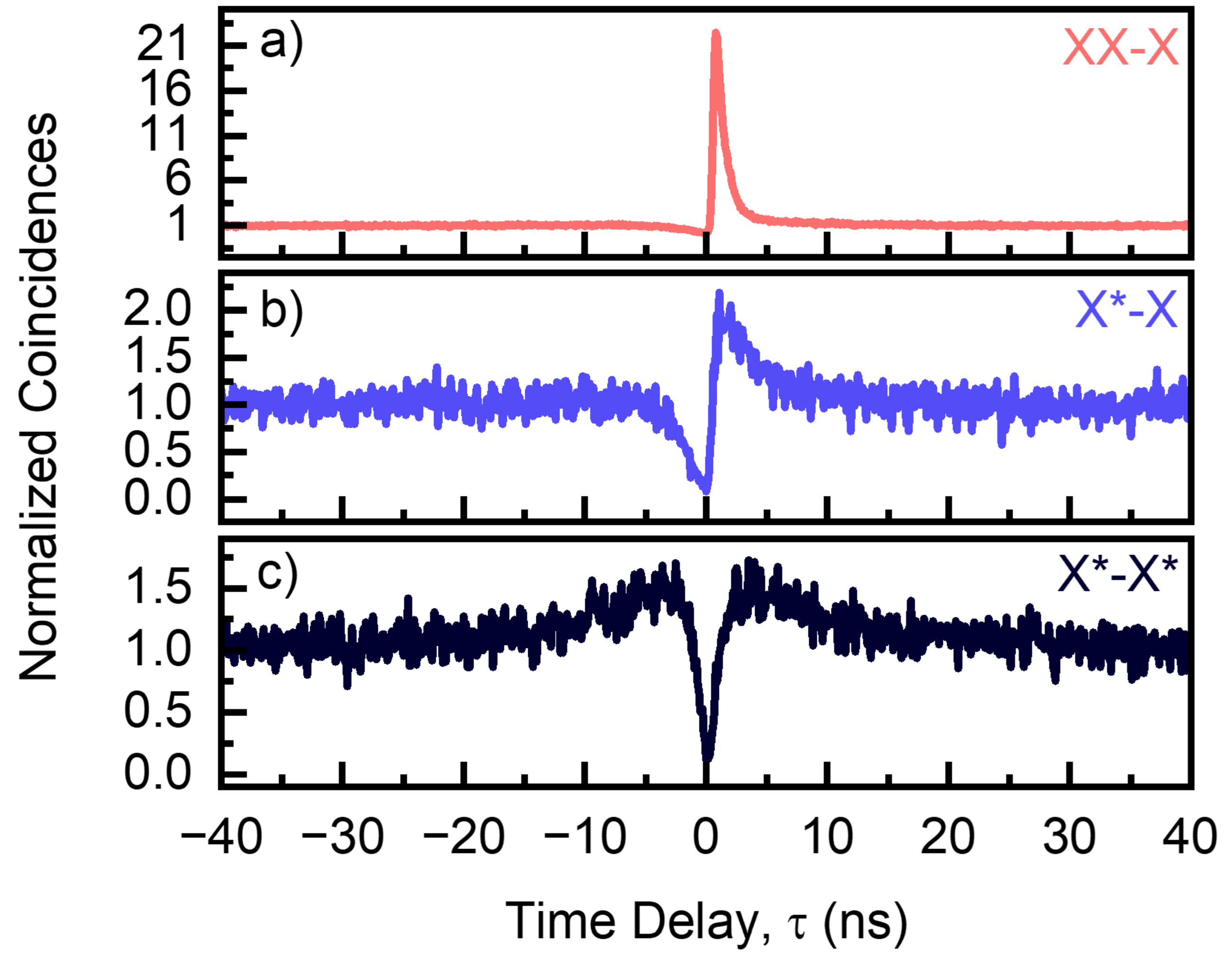}
    \end{center}
    \caption{Second-order correlation measurements from the representative QD in the main text (QD\,3). Cross-correlation measurements performed on (a) XX-X and (b) X*-X, along with (c) an auto-correlation measurement performed on X*.}
    \label{fig:Cross_Corr}
\end{figure}

In both Fig.~\ref{fig:Cross_Corr}a and b, we observe anti-bunching for small negative times, followed by asymmetric bunching for small positive times. In the case of the XX-X cross-correlation (c.f. Fig.~\ref{fig:Cross_Corr}a), the observed anti-bunching at short negative time delays is attributed to the very low probability of observing X emission just before XX emission as it would require two electron-hole pairs to be rapidly trapped after the X emission.\cite{laferriere_Systematic_2021, baier_Quantum_2006} For small positive times, we see a very pronounced bunching peak indicating a strong likelihood of detecting the emission of a X photon following the emission of an XX photon. This is characteristic of the XX-X cascade, with the exponential decay being the X lifetime. \cite{baier_Quantum_2006}

We see a similar asymmetric anti-bunching behaviour for the X-X* cross-correlation (c.f. Fig.~\ref{fig:Cross_Corr}b), however, this time with a less pronounced peak for positive time delays. This is consistent with an increased likelihood of capturing a hole to create a neutral exciton, following the emission of an X*, compared to negative times $\tau$\,$<$\,0, where observation of the X means that the dot is then empty requiring capture of three carriers to observe the X*.\cite{baier_Quantum_2006}

Finally, performing an auto-correlation measurement on the X* complex, we see bunching for both small positive and negative time delays. Indicative of blinking, it suggests that this is not the preferred charge configuration for this dot, which is often the case for charged complexes.\cite{sallen_Exciton_2009}

\subsection{Quantum Dot Linewidths}

\begin{figure}[htb]
    \begin{center}
    \includegraphics[width=\columnwidth]{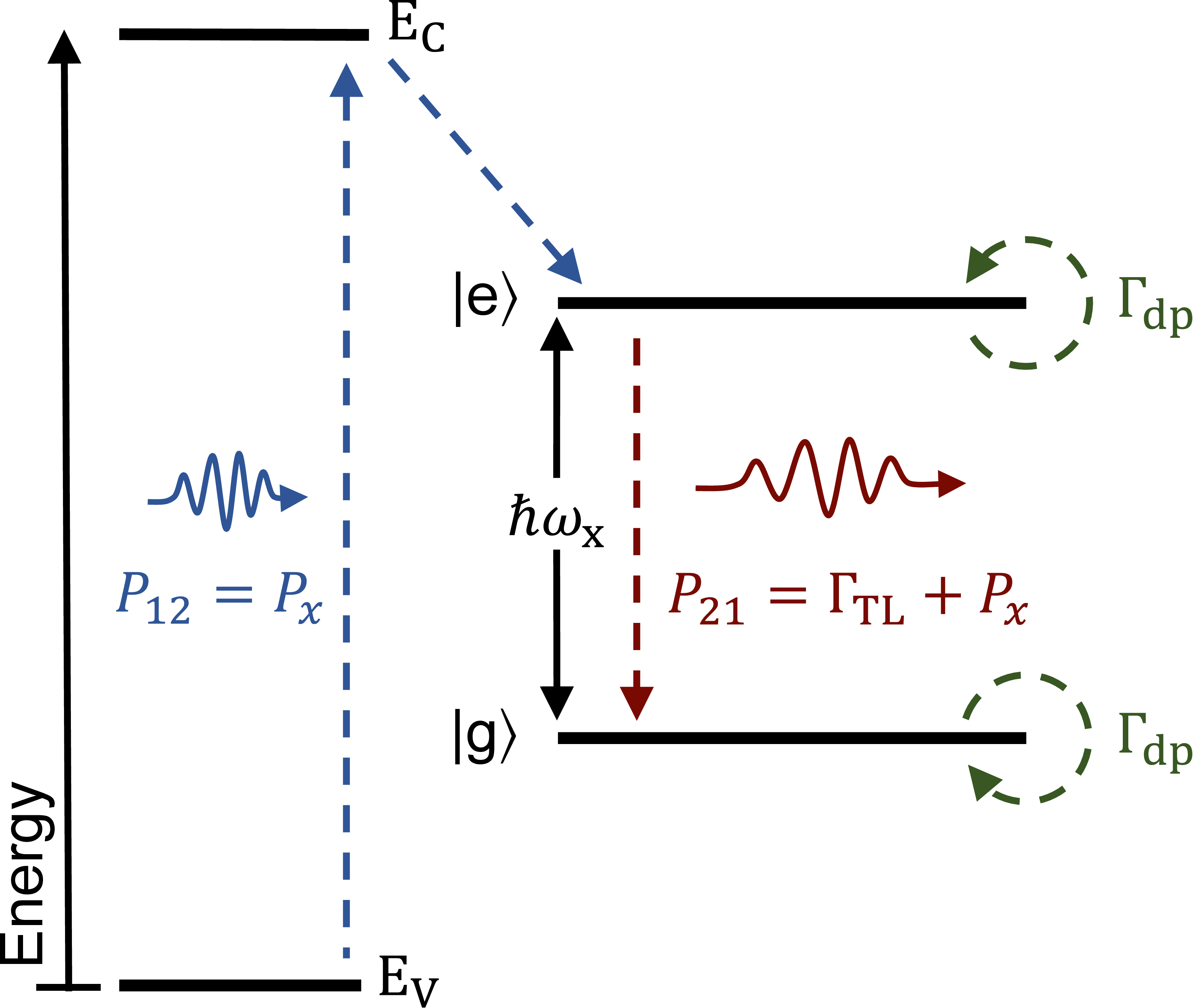}
    \end{center}
    \caption{Quantum dot two-level system. A diagram depicting the ground ($\ket{g}$) and excited ($\ket{e}$) states of a quantum dot. Under incoherent excitation, the QD is incoherently pumped with an effective pump rate $P_{12}=P_x$ that includes optical excitation of charge carriers above the bandgap which then decay into the excited state of the QD via non-radiative relaxation. While in the excited state, the QD coherence is reset at the pure dephasing rate $\Gamma_{\mathrm{dp}}$. The QD then decays back to the ground state emitting a photon with an energy $\hbar\omega_x$ at an effective decay rate $P_{21}$, set by the natural decay rate of the dot $\Gamma_{\mathrm{TL}}$ and the effective pump rate.}
    \label{fig:TLS}
\end{figure}

In the ideal case, a QD will have a linewidth set by the Fourier transform of its radiative lifetime. This transform-limited linewidth ($\Gamma_{\mathrm{TL}}$) is given by
 \begin{equation}
     \Gamma_{\mathrm{TL}} = \frac{1}{2\pi T_1}
 \end{equation}where $T_1$ is the radiative lifetime. Any broadening of this stems from environmental influences, such as uncontrolled charge fluctuations, or phonon and spin interactions \cite{anderson_Coherence_2021}. Further, the wetting layer associated with SK QDs is known to act as a reservoir for charge carriers \cite{anderson_Coherence_2021}, contributing significantly to charge noise and spectral wandering. In contrast, QDs grown with a suppressed or minimal wetting layer (as is the case for DE-grown QDs) have been shown to exhibit greater spectral stability and longer coherence times \cite{lobl_Excitons_2019, anderson_Coherence_2021}. 

Optical excitation of the QD (particularly above-band excitation) can inject charges into the solid-state matrix surrounding the QD and lead to significant increases in the linewidth for increasing excitation power\cite{Laferriere_Approaching-transform_2023}. 

To model linewidth broadening due to incoherent excitation, we follow a rigorous \textit{thermal bath model}\cite{Yao_NonlinearPhotoluminescence_2010} and treat the QD as a two level system weakly coupled to an effective thermal bath, which provides incoherent pumping via carrier relaxation from higher energy states (depicted in Fig.~\ref{fig:TLS}). The system density matrix $\rho(t)$ obeys the Born-Markov master equation

\begin{equation}
    \frac{d\rho}{dt} = \frac{-i}{\hbar}[H_s,\rho] + \mathcal{L}(\rho),
\end{equation}

\noindent with the system Hamiltonian $H_s=\hbar\omega\sigmaeg\sigmage$ and Lindblad superoperator

\begin{align}       
    \mathcal{L}(\rho) &= \frac{P_{12}}{2}(2 \sigmaeg\rho\sigmage - \sigmage\sigmaeg\rho - \rho\sigmage\sigmaeg) \nonumber \\
    &  \quad + \frac{P_{21}}{2}(2\sigmage\rho\sigmaeg - \sigmaeg\sigmage\rho - \rho\sigmaeg\sigmage) \nonumber \\
    & \quad + \frac{\Gamma_{\mathrm{dp}}}{4}(\hat\sigma_z\rho\hat\sigma_z - \rho).
\end{align}

Here, $\sigmaeg=\ket{e}\bra{g}$, $\sigmage=\ket{g}\bra{e}$, and $\hat\sigma_z=\sigmaeg\sigmage - \sigmage\sigmaeg$ are the atomic operators that describe the transitions between $\ket{e}$ and $\ket{g}$, $P_{12}$ and $P_{21}$ are the upward and downward incoherent pump rates respectively and $\Gamma_{\mathrm{dp}}$ is the pure dephasing rate. Consistent with the \textit{thermal bath model}\cite{Yao_NonlinearPhotoluminescence_2010}, we set $P_{12}=P_x$, and $P_{21} = \Gamma_{\mathrm{TL}} + P_x$, where $P_x$ is the effective pump rate of the QD due to the incoherent relaxation of electron-hole pairs from higher energy levels.

This gives us the following equations of motion

\begin{align}
    \dot{\rho}_{ee} &= -\dot{\rho}_{gg} = P_{12}\rho_{gg} - P_{21}\rho_{ee}\\
    \dot{\rho}_{eg} &= -\rho_{eg}(i\omega_x + \Gamma_2)\\
    \dot{\rho}_{ge} &= \rho_{ge}(i\omega_x - \Gamma_2)\\ \nonumber
\end{align} 

\noindent where we define $\Gamma_2 = \frac{1}{2}(\Gamma_{\mathrm{TL}} + 2P_{x} + \Gamma_{\mathrm{dp}})$ as the effective decay rate of the QD and $\omega_x$ as the transition frequency of our two-level system. Solving these differential equations, we obtain the following time dependent solutions of our system

\begin{align}
    \rho_{ee}(t) &= \rho_{ee}^{(ss)} + \left(\rho_{ee}(0) - \rho_{ee}^{(ss)}\right)e^{-(P_{12}+P_{21})t}\\
    \rho_{gg}(t) &= \rho_{gg}^{(ss)} + \left(\rho_{gg}(0) - \rho_{gg}^{(ss)}\right)e^{-(P_{12}+P_{21})t}\\
    \rho_{eg}(t) &= \rho_{eg}(0)e^{-(i\omega_x + \Gamma_{dp})t}\\
    \rho_{ge}(t) &= \rho_{ge}(0)e^{(i\omega_x - \Gamma_{dp})t}\\
\end{align}

\noindent with their corresponding steady-state behaviour (when $\dot{\rho}=0$),

\begin{align}
    \rho_{ee}^{(ss)} &= \frac{P_{12}}{P_{12}+P_{21}}\\
    \rho_{gg}^{(ss)} &= \frac{P_{21}}{P_{12}+P_{21}}\\
    \rho_{eg}^{(ss)} &= 0\\
    \rho_{ge}^{(ss)} &= 0\\
\end{align}

Now that the master equation has been solved, our task is to obtain the effective emission spectra of our system, $F(\mathrm{\omega})=\mu S(\mathrm{\omega})$. Here, $\mu$ is a collection efficiency factor that depends on the detector/collection optics, and $S(\omega)$ is the effective QD spectral response  

\begin{align}
    S(\omega) &= \frac{\Gamma_{\mathrm{TL}}}{\pi}\lim_{t \to \infty}\Re\!\int^{\infty}_0e^{i\omega \tau}\langle{\sigmaeg(t)\sigmage(t+\tau)}\rangle d\tau.
\end{align}

\noindent which is given by the Fourier transform of the two-time correlation function $\langle{\sigmaeg(t)\sigmage(t+\tau)}\rangle$. To evaluate this, we will use the quantum regression theorem \cite{Steck_QuantumAtomOptics_2007}, which states that two-time correlation functions obey the same equations of motion as one-time averages, such that

\begin{equation}
    \lim_{t \to\inf}\langle{\sigmaeg(t)\sigmage(t+\tau)}\rangle\ = \mathrm{Tr}[\sigmage\Lambda(\tau)],
\end{equation}

\noindent Here we define $\Lambda(\tau)$ as our two-time operator, which satisfies the same equations of motion as $\rho$(t) and has initial conditions $\Lambda(0) = \rho(t\rightarrow\infty)\sigmaeg$. In doing so, we obtain

\begin{equation}
    \langle{\sigmaeg(t)\sigmage(t+\tau)}\rangle\ = \Lambda_{eg}(\tau) = \rho_{ee}^{(ss)}e^{-(i\omega_x + \Gamma_2)}
\end{equation}

\noindent and the effective QD spectral response can be rewritten as

\begin{align}
    F(\omega) &= \frac{\mu\Gamma_{\mathrm{TL}}}{\pi}\Re\!\int^{\infty}_0\rho_{ee}^{(ss)}e^{-(\Gamma_2-i(\omega-\omega_x)) \tau}d\tau \nonumber \\
    &= \frac{\mu\Gamma_{\mathrm{TL}}}{\pi}\rho_{ee}^{(ss)}\frac{\Gamma_2}{\Gamma_2^2 + (\omega-\omega_x)^2},
\end{align}
 
 \noindent where we see that $\Gamma_2$ is the half-width at half-maximum (HWHM) linewidth, and is related to the full-width at half-maximum (FWHM) linewidth (in linear frequency) as $\Gamma_{\mathrm{meas}} = \Gamma_2/\pi$. Notably, this includes contributions due to the radiative decay rate, pure dephasing rate and importantly the effective pump rate of the dot. To obtain the emission intensity of the dot $I(P_x)$, we integrate over all frequencies to obtain

\begin{align}
    I(P_x)& = \frac{\mu\Gamma_{\mathrm{TL}}}{\pi}\rho_{ee}^{(ss)}\int^{\infty}_{-\infty}\frac{1}{\Gamma_2^2 + (\omega-\omega_x)^2}d\omega \nonumber \\\
    &= \mu\Gamma_{\mathrm{TL}}\rho_{ee}^{(ss)}
\end{align}

Our goal now is to express this in terms of the excitation pump power $\mathrm{P}$, instead of $P_x$, in order to quantify the amount of pure dephasing present in each dot. Once known, we can solve for the effective decay rate in the low-power limit, given by $\Gamma_2(\mathrm{P}\rightarrow0) = \frac{1}{2}(\Gamma_{\mathrm{TL}} + \Gamma_{\mathrm{dp}})$. For an ideal two-level system, the effective pump rate is directly proportional to the excitation power such that $P_x\propto \mathrm{P}$. However, real QDs often deviate from this due to various carrier capture mechanisms such as inefficient carrier relaxation, or in the case of singly charged excitons\cite{laferriere_Systematic_2021} (which we study in depth in this work). Therefore, for full generality, we will assume that $P_x(\mathrm{P}) = \alpha\mathrm{P}^n$. 

Considering that for the \textit{thermal bath model}, when at steady state, P$=\mathrm{P_{sat}}$, and $P_x \approx \Gamma_{\mathrm{TL}}/2$, we can rearrange to solve for $\alpha = \Gamma_{\mathrm{TL}}/2\mathrm{P_{sat}}^n$, giving us

\begin{equation}
    P_x(\mathrm{P}) = \frac{\Gamma_{\mathrm{TL}}}{2}\left({\frac{\mathrm{P}}{\mathrm{P_{sat}}}}\right)^n.
\end{equation}

\noindent Substituting this back into our expression for the emission intensity, we obtain an expression explicitly as a function of excitation pump power,

\begin{equation}
    I(\mathrm{P}) = \mathrm{I_{sat}}\frac{\mathrm{P}^n}{\mathrm{P_{sat}} + \mathrm{P}^n},
\end{equation}
 where $\mathrm{I_{sat}} = \mu\Gamma_{\mathrm{TL}}$/2 is the emission intensity at saturation. 

Finally, substituting $P_x(\mathrm{P})$ into $\Gamma_2$ and converting to linear frequency, we obtain the following expression for the FWHM spectral linewidth as a function of excitation power

\begin{equation}
    \Gamma_{\mathrm{meas}}(\mathrm{P})  = \frac{1}{2\pi T_1}\left[1 + \left(\frac{\mathrm{P}}{\mathrm{P_{sat}}}\right)^{n}\right] + \frac{\Gamma_{\mathrm{dp}}}{2\pi}
\end{equation}

\noindent where $T_1 = \Gamma_{\mathrm{TL}}^{-1}$ is the radiative decay lifetime of the QD. 

As an example of how we use this to estimate the intrinsic linewidth in the low-power limit $(\mathrm{P}\rightarrow0)$, we consider the X* linewidth of the representative quantum dot in the main text (QD\,3). Reaching saturation at 16.4$\pm$2.7\,$\mu$W with a rate of $n=1.09\pm0.04$, and lifetime of 1.57$\pm$0.02\,ns, we first determine the amount of pure dephasing given a measured linewidth of $\Gamma_{\mathrm{meas}}=1.8\pm0.4$\,GHz at a power of $1.2\pm0.2\mathrm{P_{sat}}$ and then estimate the extracted linewidth at zero excitation power, 2$\Gamma_2(\mathrm{P\rightarrow0)}$. Because the etalon used to measure the linewidths has a Lorentzian instrument response function (IRF) of 175$\pm$25\,MHz, the resulting lineshape is a convolution between the Lorentzian QD linewidth and the IRF, with a spectral width set by the sum of the two. By subtracting off the IRF of the etalon, we recover the corrected QD linewidth, $\Gamma_{\mathrm{meas}}^{(IRF)}$. Using Eq.\,S.24,

\begin{align}
    \frac{\Gamma_{\mathrm{dp}}}{2\pi} &= \Gamma_{\mathrm{meas}}^{(IRF)}(\mathrm{P}) - \frac{1}{2\pi T_1}\left[1+2\left(\frac{\mathrm{P}}{\mathrm{P_{sat}}}\right)^n \right] \nonumber\\
    & = 1495\pm278\,\mathrm{MHz}
\end{align}

\begin{align}
    \frac{\Gamma_2(\mathrm{P\rightarrow0})}{\pi} &= \frac{1}{2\pi T_1} + \frac{\Gamma_{\mathrm{dp}}}{2\pi} \nonumber\\
    &= 1597\pm279\,\mathrm{MHz}
\end{align}

We note that depending on the value of n (which would typically be $\sim1$), Eq.\,S.24 is roughly linearly proportional to excitation power, however, deviations from this expected behaviour has been observed with incoherently pumped QDs, particularly at excitation powers several times P$_{\mathrm{sat}}$ \cite{Laferriere_Approaching-transform_2023}. One possible explanation for this could be due to additional broadening mechanisms related to phonons, which we do not consider in our model. Particularly with incoherent excitation, large pump powers may stimulate the material lattice surrounding the quantum dot, generating additional phonons that lead to pronounced QD dephasing.

\subsection{Summary on Telecom-Wavelength InAs/InP QD Linewidths}

\begin{table*}[t]
    \centering
    \caption{Comparison of reported C-band emitting InAs/InP QD linewidths.\protect\tabularnote{GM: growth method; CBE: chemical-beam epitaxy; MOVPE: metalorganic vapour-phase epitaxy; MBE: molecular-beam epitaxy; SK; Stranski-Krastanov, DE: droplet-epitaxy; AB: above-band; CBR: circular Bragg grating; PhCM: photonic crystal membrane; DBR: distributed Bragg reflector; TPE: two-photon resonant; RRS: resonant Rayleigh scattering}}
    \resizebox{\textwidth}{!}{%
    \begin{NiceTabular}{cccccccc} 
        \toprule
        \textbf{GM} & \textbf{QD Type} & \textbf{Device} & \textbf{Exc. Scheme} &
        \textbf{Lifetime} & \textbf{$\Gamma$} & \textbf{$\Gamma / \Gamma_{TL}$} & \textbf{Ref}\\
        \midrule
        \multirow{2}{*}{CBE} & \multirow{2}{*}{SK} & \multirow{2}{*}{bottom mirror} & \multirow{2}{*}{AB} & 1.64$\pm$0.02\,ns\tabularnote{Average lifetime} & 1384$\pm$843\,MHz\tabularnote{Average linewidth} & 12.1$\pm$6.7$\times$ & \multirow{2}{*}{this work} \\
        & & & & 1.64$\pm$0.02\,ns\tabularnote{Average lifetime} & 272$\pm$171\,MHz\tabularnote{Best linewidth} & 2.8$\pm$1.8$\times$ & \\
        \cmidrule(lr){1-8}
        MOVPE & SK & bottom mirror & TPE & 0.34\,ns \tabularnote{Bi-exciton lifetime and corresponding linewidth} & 11.3\,GHz & 24.1$\times$ & Vajner et al. 2024 \cite{vajner_-demand_2024}\\
        \cmidrule(lr){1-8}
        MOVPE & SK & bottom mirror + CBR & AB & 0.53\,ns \tabularnote{Purcell enhanced lifetime} & 21.2\,GHz\tabularnote{Spectrometer resolution limited measurement} & 70.7$\times$ & Holewa et al. 2024 \cite{holewa_high-throughput_2024}\\
        \cmidrule(lr){1-8}
        \multirow{2}{*}{MOVPE} & \multirow{2}{*}{DE} & \multirow{2}{*}{PhCM} & \multirow{2}{*}{AB} & 0.34\,ns \tabularnote{Purcell enhanced lifetime} & \multirow{2}{*}{4.8\,GHz} & 10.3$\times$ & \multirow{2}{*}{Phillips et al. 2024 \cite{phillips_purcell-enhanced_2024}}\\
        & & & & 1.79\,ns & & 54.0$\times$ & \\
        \cmidrule(lr){1-8}
        MBE & SK & nanobeam & AB & 1.87\,ns & 8.36\,GHz\tabularnote{Spectrometer resolution limited measurement} & 98.4$\times$ & Rahaman et al. 2024 \cite{rahaman_efficient_2024}\\
        \cmidrule(lr){1-8}
        MOVPE & DE & DBR cavity + gates & RRS & ~1\,ns\tabularnote{Average lifetime} & 653\,MHz & 4.1$\times$ & Wells et al. 2023 \cite{wells_coherent_2023}\\
        \cmidrule(lr){1-8}
        MBE & SK & DBR & AB & 1.79\,ns & 9.2\,GHz\tabularnote{Spectrometer resolution limited measurement} & 103.4$\times$ & Musial et al. 2020 \cite{musial_high-purity_2020}\\
        \bottomrule
    \end{NiceTabular}}
    \label{Linewidth_summary_table}
\end{table*}

\begin{figure}[htb]
    \begin{center}
    \includegraphics[width=\columnwidth]{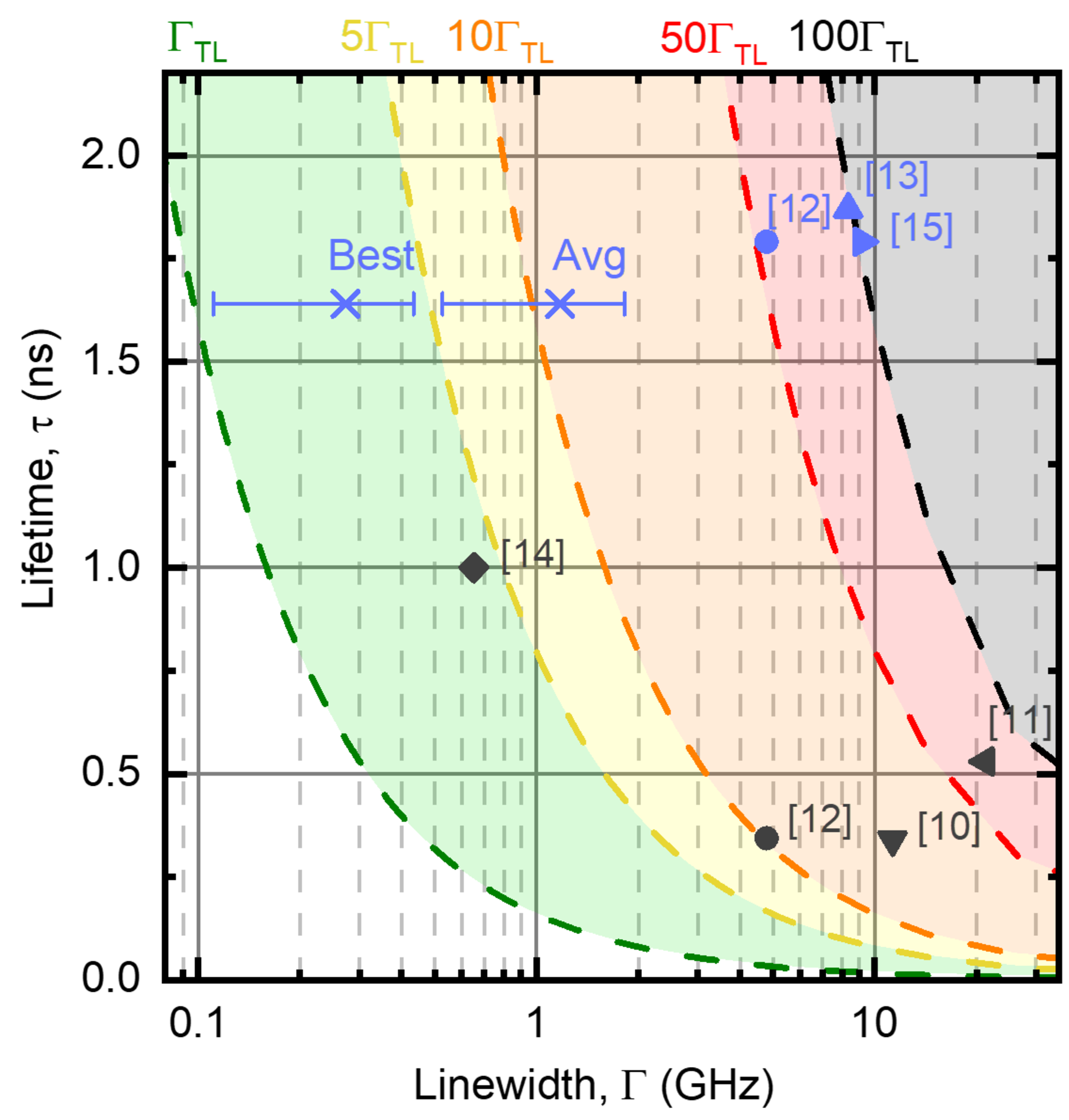}
    \end{center}
    \caption{Reported linewidths from InAs/InP QDs emitting in the telecom C-band. Here, light blue symbols correspond to linewidth measurements obtained under above-band excitation on QDs without any Purcell enhancement, whereas black symbols correspond to measurements obtained under sophisticated excitation schemes, photonic cavities, and/or electrical gating. Regions for 1-5\,$\Gamma_{\mathrm{TL}}$ (green), 5-10\,$\Gamma_{\mathrm{TL}}$ (yellow), 10-50\,$\Gamma_{\mathrm{TL}}$ (orange), 50-100\,$\Gamma_{\mathrm{TL}}$ (red) and $>$100\,$\Gamma_{\mathrm{TL}}$ (dark grey) are marked by dashed lines and shaded regions, coloured respectively. For additional details regarding each data point, see Table~\ref {Linewidth_summary_table}.}
    \label{fig:linewidth_summary}
\end{figure}

In this section, we summarize recent reports on InAs/InP QD linewidth measurements over the last six years. In preparing this summary, several considerations were taken into account, namely what growth system/method, along with the specific growth scheme was used to create the dots, whether any additional nanophotonic devices were used to enhance emission rates via lifetime enhancements, and finally, what excitation method was used to perform the measurement. This is summarized in Table.~\ref{Linewidth_summary_table} and in Fig.~\ref{fig:linewidth_summary}

We begin by comparing our results to those performed under similar conditions, specifically non-resonant excitation and no lifetime enhancements.\cite{phillips_purcell-enhanced_2024, musial_high-purity_2020, rahaman_efficient_2024} Here we see in all cases linewidths of greater than 50\,$\Gamma_{\mathrm{TL}}$, however we note that in two of these cases the authors state measurements were limited by the resolution of their spectrometer, so these values are likely an upper bound\cite{rahaman_efficient_2024, musial_high-purity_2020}. As stated in the main text, planar quantum dots are compatible with a number of photonic devices that can enable lifetime enhancement via the Purcell effect. As is the case with Phillips et al.\cite{phillips_purcell-enhanced_2024}, assuming a consistent linewidth of 4.8\,GHz, by using a photonic crystal membrane to give a Purcell enhancement of $\sim$5, they come close to 10\,$\Gamma_{\mathrm{TL}}$, or a factor of $\sim$5$\times$ improvement. 

Following this, researchers using coherent excitation schemes which more efficiently excite the QDs while minimizing the number of carriers ejected into the surrounding material report linewidths of only 24\,$\Gamma_{\mathrm{TL}}$ (measured on the XX)\cite{vajner_-demand_2024}. Based on previous results, we can assume a 2$\times$ linewidth improvement here. Finally, applying both resonant excitation schemes and electrical stabilization of the surrounding charge environment, Wells et al.\cite{wells_coherent_2023} report a measured linewidth of only 4\,$\Gamma_{\mathrm{TL}}$, roughly a factor of 13$\times$ in linewidth improvement over non-resonant excitation reports. 

\onecolumn
\bibliography{main}